\documentstyle[preprint,aps,epsfig]{revtex}
\tightenlines
\begin{document}
\draft

\title{Role of the nonperturbative input in QCD resummed Drell-Yan
$Q_T$-distributions}
\author{Jianwei Qiu and Xiaofei Zhang}
\address{Department of Physics and Astronomy,
         Iowa State University \\
         Ames, Iowa 50011, USA}

\date{February 16, 2001}
\maketitle
\begin{abstract}
We analyze the role of the nonperturbative input in the Collins,
Soper, and Sterman (CSS)'s $b$-space QCD resummation formalism for 
Drell-Yan transverse momentum ($Q_T$) distributions, and investigate
the predictive power of the CSS formalism.  We find that the
predictive power of the CSS formalism has a strong dependence
on the collision energy $\sqrt{S}$ in addition to its well-known $Q^2$
dependence, and the $\sqrt{S}$ dependence improves the predictive
power at collider energies.  We show that a reliable extrapolation
from perturbatively resummed $b$-space distributions to the
nonperturbative large $b$ region is necessary to ensure the correct
$Q_T$ distributions.  By adding power corrections to the
renormalization group equations in the CSS formalism, we derive a new
extrapolation formalism.  We demonstrate that at collider energies,
the CSS resummation formalism plus our extrapolation has an excellent
predictive power for $W$ and $Z$ production at all transverse momenta
$Q_T\le Q$.  We also show that the $b$-space resummed $Q_T$
distributions provide a good description of Drell-Yan data at fixed
target energies.    
\end{abstract}
\vspace{0.2in}

\pacs{PACS number(s): 12.38.Cy, 12.38.Qk, 14.70.-e}

\section{Introduction}

The perturbation theory of Quantum Chromodynamics (QCD) has been very
successful in interpreting and predicting high energy scattering
processes.  With new data from Fermilab Run II and the LHC in the near
future, we expect to test QCD to a new level of accuracy, and also
expect that a better understanding of 
QCD will underpin precision tests of the Electroweak interactions and
particle searches beyond the Standard Model \cite{QCD-rpt}.  
As pointed out in Ref.~\cite{QCD-rpt}, the description of vector and
scalar boson production properties, in particular their transverse
momentum ($Q_T$) distribution, is likely to be one of the most
intensively investigated topics at both Fermilab and the LHC,
especially in the context of Higgs searches.  It is the purpose of
this paper to investigate the predictive power of QCD perturbation
theory for the transverse momentum distributions of vector and scalar
boson production in hadronic collisions. 

The production of vector bosons ($V=\gamma^*, W^{\pm}$, and $Z$) with 
invariant mass $Q$ at large and small $Q_T$ in hadronic collisions has
been extensively studied \cite{QCD-rpt}.  When $Q_T \sim Q$, 
effectively, there is 
only one hard momentum scale in the production.  Therefore, we expect
the fixed-order perturbative calculations in power series of
$\alpha_s$ to be reliable \cite{DY-fix}.  In this paper, we will
concentrate on the production of vector bosons at small transverse 
momenta: $Q_T\leq Q$, where the bulk of the data is.  The small $Q_T$
region also corresponds to a phase space that is most relevant to the
hadronic Higgs production.   

When $Q_T \ll Q$, the $Q_T$ distributions calculated order-by-order in  
$\alpha_s$ in the conventional fixed-order perturbation theory
receive a large logarithm, $\ln(Q^2/Q_T^2)$, at every power of
$\alpha_s$.  Even at the leading order in $\alpha_s$, the cross
section $d\sigma/dQ^2\,dQ^2_T$ contains a term proportional to
$(\alpha_s/Q_T^2)\ln(Q^2/Q_T^2)$ coming from the partonic subprocess:  
$q+\bar{q} \rightarrow V(\gamma^*,W/Z)+g$.  Beyond the leading order,
we can actually get two powers of the logarithm for
every power of $\alpha_s$, due to soft and collinear gluons emitted by
the incoming partons.  
Therefore, at sufficiently small $Q_T$, the convergence of the
conventional perturbative expansion in powers of $\alpha_s$ is
impaired, and the logarithms must be resummed.   

Resummation of the large logarithms in QCD can be carried out either
in $Q_T$-space directly, or in the impact parameter, $b$-space, which
is a Fourier conjugate of the $Q_T$-space.  It was first shown by
Dokshitzer, Diakonov and Troyan (DDT) that in the double leading
logarithm approximation (DDLA), which corresponds to the phase space
where the radiated gluons are both soft and collinear with strong
ordering in their transverse momenta, the dominant contributions in
the small $Q_T$ region can be resummed into a Sudakov form factor
\cite{DDT-qt}.  However, the strong ordering in transverse momenta in
the DDLA overly constrains the phase space of the emitted soft gluons
and ignores the overall momentum conservation.  As a result, the DDT
resummation formalism unphysically suppresses the $Q_T$ distributions
at small $Q_T$ \cite{QCD-rpt}.  By imposing the transverse momentum
conservation without assuming the strong ordering in transverse
momenta of radiated gluons, Parisi and Petronzio introduced the
$b$-space resummation method which allows resummation of some
subleading logarithms \cite{PP-b}. By using the renormalization group
equation technique, Collins and Soper improved the $b$-space
resummation to resum all logarithms as singular as,
$\ln^m(Q^2/Q_T^2)/Q_T^2$, as $Q_T\rightarrow 0$ \cite{CS-b}.  In the
frame work of this renormalization group improved $b$-space
resummation, Collins, Soper, and  Sterman (CSS) derived a formalism
for the transverse momentum distributions of vector boson production
in hadronic collisions \cite{CSS-W}.  This formalism, which is often
called CSS formalism, can be also applied to hadronic productions of
Higgs bosons \cite{DY-higgs}.  

For the Drell-Yan vector boson production in hadronic collisions
between hadrons $A$ and $B$, $A(P_A)+B(P_B)\rightarrow V(Q)+X$ with
$V=\gamma^*,W^{\pm}$, and $Z$, the CSS resummation formalism has the
following generic form \cite{CSS-W} 
\begin{equation}
\frac{d\sigma(h_A+h_B\rightarrow V+X)}{dQ^2\, dy\, dQ_T^2} =
\frac{1}{(2\pi)^2}\int d^2b\, e^{i\vec{Q}_T\cdot \vec{b}}\,
\tilde{W}(b,Q,x_A,x_B) + Y(Q_T,Q,x_A,x_B)\, ,
\label{css-gen}
\end{equation}
where $x_A= e^y\, Q/\sqrt{S}$ and $x_B= e^{-y}\, Q/\sqrt{S}$ with the
rapidity $y$ and collision energy $\sqrt{S}$.  In Eq.~(\ref{css-gen}),
the $\tilde{W}$ term dominates the $Q_T$ distributions when $Q_T\ll
Q$, and the $Y$ term gives corrections that are negligible for small
$Q_T$, but becomes important when $Q_T\sim Q$.  In the CSS formalism,
the $\tilde{W}$ has the following form \cite{CSS-W}
\begin{equation}
\tilde{W}(b,Q,x_A,x_B) = \sum_{ij} 
  \tilde{W}_{ij}(b,Q,x_A,x_B)\, \sigma_{ij\rightarrow V}(Q)\, ,
\label{css-W-ij}
\end{equation}
where $\sigma_{ij\rightarrow V}(Q)$ is the lowest order cross section
for a pair of quark and antiquark of invariant mass $Q$ to annihilate
into a vector boson $V$, and the $\sum_{ij}$ run over all
possible quark and antiquark flavors that can annihilate into the
vector boson at the Born level.  In Eq.~(\ref{css-W-ij}), the
$\tilde{W}_{ij}(b,Q,x_A,x_B)$ is an effective flux to have partons of
flavor $i$ and $j$ from the respective hadron $A$ and $B$, and it
has the following form \cite{CSS-W}   
\begin{equation}
\tilde{W}_{ij}(b,Q,x_A,x_B) = 
{\rm e}^{S(b,Q)}\, \tilde{W}_{ij}(b,c/b,x_A,x_B)\, ,
\label{css-W-sol}
\end{equation}
where $S(b,Q)$ will be specified later and $c$ is a constant of order
one \cite{CSS-W,QZ1}.  The $\tilde{W}_{ij}(b,c/b,x_A,x_B)$ in 
Eq.~(\ref{css-W-sol}) depends on only one momentum scale, $1/b$, and
is perturbatively calculable as long as $1/b$ is large
enough.  All large logarithms from $\ln(1/b^2)$ to $\ln(Q^2)$ in
$\tilde{W}_{ij}(b,Q,x_A,x_B)$ are completely resummed into the
exponential factor $\exp[S(b,Q)]$.  

Since the perturbatively resummed $\tilde{W}_{ij}(b,Q,x_A,x_B)$ in
Eq.~(\ref{css-W-sol}) is only reliable for the small $b$ region, an
extrapolation to the large $b$ region is necessary in order to
complete the Fourier transform in Eq.~(\ref{css-gen}).
In the CSS formalism, a variable $b_*$ 
and a nonperturbative function $F^{NP}(b,Q,x_A,x_B)$
were introduced \cite{CSS-W}, 
\begin{equation}
\tilde{W}^{\rm CSS}(b,Q,x_A,x_B) \equiv 
\tilde{W}(b_*,Q,x_A,x_B)\,
F^{NP}(b,Q,x_A,x_B)\, ,
\label{css-W-b}
\end{equation}
where $b_*=b/\sqrt{1+(b/b_{max})^2} < b_{max} = 0.5$~GeV$^{-1}$, and
$F^{NP}$ has a Gaussian-like dependence on $b$, $F^{NP} \sim
\exp(-\kappa b^2)$ and the parameter $\kappa$ has some dependence on
$Q^2$, $x_A$, and $x_B$.  

The predictive power of the CSS $b$-space resummation formalism relies
on the following criterions: (1) the Fourier 
transform (or $b$-integration) in Eq.~(\ref{css-gen}) is dominated by
the perturbative (or small $b$) region; and (2) the nonperturbative
input $F^{NP}$ has a derived $Q$ dependence when $Q$ is large.  By
using data at some values of $Q$ to fix the nonperturbative
$b$-dependence of $F^{NP}$, the CSS formalism predicts the $Q_T$
distributions 
at different values of $Q$.  Existing data are not inconsistent with
such a form \cite{Davis,AK,LY,Ellis-1,LBLY}.  However, improvements
are definitely needed for the precision tests of the theory
\cite{QCD-rpt,Ellis-2}. 

Although the $b$-space resummation formalism has been successful in
interpreting existing data, it was argued \cite{QCD-rpt,Ellis-2}
that the formalism has many drawbacks associated with working in
impact parameter space.  As listed in Ref.~\cite{QCD-rpt}, the
first is the difficulty of matching the resummed and fixed-order
predictions; and the second is to know the quantitative difference
between the prediction and the fitting because of the introduction of
a nonperturbative $F^{NP}$.  In viewing of these
difficulties, major efforts have been devoted to resum the large
logarithms directly in $Q_T$-space \cite{QCD-rpt,Ellis-2}.

In this paper, we argue and demonstrate that both of these
drawbacks can be overcomed \cite{QZ1}.  Since there is no preferred
transverse direction, the two-dimensional Fourier transform in
Eq.~(\ref{css-gen}) can be reduced into an one-dimensional integration
over $b=|\vec{b}|$ weighted by the Bessel function $J_0(Q_Tb)$
\cite{PP-b,CSS-W}.  We find that by using an integral form for the
Bessel function, the $b$-space resummation formalism
works smoothly for all $Q_T \leq Q$.  Because of the smooth resummed
$Q_T$ distributions, the transition (or switch over) to the fixed
order perturbative calculations at large $Q_T$ becomes less ambiguous
\cite{Ellis-2}. 

In addition, we find that the predictive power of the $b$-space
resummation formalism strongly depends on the 
collision energy $\sqrt{S}$ in addition to its well-known $Q^2$
dependence.  Because of the steep evolution of
parton distributions at small $x$, the $\sqrt{S}$ dependence of the
$\tilde{W}$ in Eq.~(\ref{css-gen}) significantly improves the
predictive power of the $b$-space resummation formalism at collider
energies.  We demonstrate that the $b$-space resummation formalism
has an excellent predictive power for the Drell-Yan $Q_T$
distributions as long as $Q^2$ is large and/or $x_A$ and $x_B$ 
are small.

To quantitatively separate the QCD prediction from the parameter
fitting, we derive a new functional form in Eq.~(\ref{qz-W-sol-m}) to
extrapolate the perturbatively calculated $b$-space distribution
$\tilde{W}(b,Q,x_A,x_B)$ to the large $b$ region.  Our extrapolation
is derived by adding power corrections to the evolution and
renormalization group equations in the CSS resummation formalism.
Our extrapolation preserves the predictive power of perturbative
calculations at small $b$, while it provides clear physical
interpretations for all $b$-dependence in the large $b$
region.  We find that the CSS $b$-space resummation formalism plus our 
extrapolation gives an excellent description of the data on the 
Drell-Yan $Q_T$ distributions at both collider and fixed target
energies.  

The rest of this paper is organized as follows.  In Sec.~\ref{sec2},
we briefly review the CSS $b$-space resummation formalism for the
Drell-Yan transverse momentum distributions.  We show that  
the predictive power of the $b$-space resummation formalism has a
significant $\sqrt{S}$ dependence.  In Sec.~\ref{sec3}, we 
quantitatively analyze the role of the nonperturbative input at large
$b$ in the $b$-space resummation formalism.  By adding power
corrections to the evolution equation of $\tilde{W}_{ij}$ and
power corrections to the renormalization group equations of
corresponding evolution kernels, we derive a new functional form to
extrapolate the 
perturbatively calculated $\tilde{W}_{ij}$ to the large $b$ region.
This new functional form clearly separates the perturbative
predictions in the small $b$ region from the nonperturbative physics
at large $b$.  Finally, in Sec.~\ref{sec4}, we numerically compare the 
$b$-space resummed $Q_T$ distributions with experimental data.  Our
conclusions are also given in Sec.~\ref{sec4}.


\section{The Collins-Soper-Sterman Formalism}
\label{sec2}

In this section, we briefly review the CSS $b$-space resummation
formalism for the Drell-Yan transverse momentum distributions. We show
that the predictive power of the $b$-space resummation formalism has a
strong dependence on the collision energy $\sqrt{S}$ in addition to
its well-known $Q^2$ dependence.  We demonstrate that the $\sqrt{S}$
dependence significantly improves the predictive power of the
$b$-space resummation formalism at collider energies.

It was shown \cite{CSS-W} that for $b\ll 1/\Lambda_{\rm QCD}$, the
$\tilde{W}(b,Q,x_A,x_B)$ is directly related to the singular parts of
the Drell-Yan $Q_T$ distribution as $Q_T \rightarrow 0$. More
precisely, the $\tilde{W}(b,Q,x_A,x_B)$ includes all singular terms
like $\delta^2(\vec{Q}_T)$ and $[\ln^m(Q^2/Q_T^2)/Q_T^2]_{\rm reg}$
with $m\ge 0$.   The terms that are less singular as
$Q_T\rightarrow 0$ are included in the $Y$ term in
Eq.~(\ref{css-gen}).  The QCD resummation of the large logarithms in  
the CSS formalism is achieved by solving the evolution equation
for the $\tilde{W}_{ij}$ \cite{CSS-W}  
\begin{equation}
\frac{\partial}{\partial\ln Q^2} \tilde{W}_{ij}(b,Q,x_A,x_B)
= \left[ K(b\mu,\alpha_s(\mu)) + G(Q/\mu,\alpha_s(\mu)) \right]
\tilde{W}_{ij}(b,Q,x_A,x_B)\, ,
\label{css-W-evo}
\end{equation}
and corresponding renormalization group equations for the kernels $K$
and $G$,
\begin{eqnarray}
\frac{\partial}{\partial\ln\mu^2} K(b\mu,\alpha_s(\mu))
&=& -\frac{1}{2} \gamma_K(\alpha_s(\mu)) \, ,
\label{css-K-rg} \\
\frac{\partial}{\partial\ln\mu^2} G(Q/\mu,\alpha_s(\mu))
&=& \frac{1}{2} \gamma_K(\alpha_s(\mu)) \, .
\label{css-G-rg}
\end{eqnarray}
The anomalous dimension $\gamma_K(\alpha_s(\mu))=\sum_{n=1} 
\gamma_K^{(n)} (\alpha_s(\mu)/\pi)^n$ in Eqs.~(\ref{css-K-rg}) and 
(\ref{css-G-rg}) are perturbatively calculable \cite{CSS-W}.  The
renormalization group equations for the $K$ and $G$ in
Eqs.~(\ref{css-K-rg}) and (\ref{css-G-rg}) ensure the correct 
renormalization scale dependence, $d/d\ln\mu^2 [ 
\tilde{W}(b,Q,x_A,x_B)] = 0$.  The solution given in
Eq.~(\ref{css-W-sol}) corresponds to  
solving the evolution equation in Eq.~(\ref{css-W-evo}) from 
$\ln(c^2/b^2)$ to $\ln(Q^2)$, and solving the renormalization group
equations in Eqs.~(\ref{css-K-rg}) and (\ref{css-G-rg}) from 
$\ln(c^2/b^2)$ to $\ln(\mu^2)$ and from $\ln(Q^2)$ to $\ln(\mu^2)$, 
respectively.  

Integrating Eq.~(\ref{css-K-rg}) over $\ln(\mu^2)$ from $\ln(c^2/b^2)$
to $\ln(\mu^2)$, and Eq.~(\ref{css-G-rg}) from $\ln(Q^2)$ to
$\ln(\mu^2)$, one derives 
\begin{equation}
K(b\mu,\alpha_s(\mu)) + G(Q/\mu,\alpha_s(\mu)) 
= -\int_{c^2/b^2}^{Q^2}\, 
  \frac{d\bar{\mu}^2}{\bar{\mu}^2}\, 
  A(\alpha_s(\bar{\mu}))
- B(\alpha_s(Q))\, ,
\label{css-KG}
\end{equation}
where $A$ is a function of $\gamma_K(\alpha_s(\bar{\mu}))$ and
$K(c,\alpha_s(\bar{\mu}))$ while $B$ depends on both
$K(c,\alpha_s(Q))$ and $G(1,\alpha_s(Q))$. 
The functions $A$ and $B$ do not have large logarithms and have
perturbative expansions $A=\sum_{n=1} A^{(n)}(\alpha_s/\pi)^n$ and
$B=\sum_{n=1} B^{(n)}(\alpha_s/\pi)^n$, respectively.
The first two coefficients in the perturbative expansions are known
\cite{CSS-W,Davis}
\begin{eqnarray}
A^{(1)} &=& C_F\ ,
\nonumber\\
A^{(2)} &=& \frac{C_F}{2} \left[
           N\left(\frac{67}{18}-\frac{\pi^2}{6}\right)
           -\frac{10}{9}T_R\, n_f \right]\, ,
\nonumber\\
B^{(1)} &=& -\frac{3}{2}C_F\, ,
\nonumber\\
B^{(2)} &=& \left(\frac{C_F}{2}\right)^2 
            \left[\pi^2 -\frac{3}{4}-12\zeta(3)\right]
          + \frac{C_F}{2}\, N
            \left[\frac{11}{18}\pi^2-\frac{193}{24} + 3\zeta(3)
                  \right]
\nonumber\\
        &+& \frac{C_F}{2}T_R\, n_f 
            \left[\frac{17}{6}-\frac{2}{9}\pi^2\right]\, ,
\label{css-AB}
\end{eqnarray}
where $N=3$ for SU(3) color, $C_F=(N^2-1)/2N=4/3$, $T_R=1/2$, and
$n_F$ is the number of active quark flavors.  The functions $A$ and
$B$ given in Eq.~(\ref{css-AB}) are derived from the general
expressions in Ref.~\cite{CSS-W} with the following choices for the
renormalization constants: $C_1=c=2{\rm e}^{-\gamma_E}$ and $C_2=1$,
where $\gamma_E\approx 0.577$ is the Euler's constant.

Substituting Eq.~(\ref{css-KG}) into Eq.~(\ref{css-W-evo}), and
integrating over $\ln(Q^2)$ from $\ln(c^2/b^2)$ to $\ln(Q^2)$, one
obtains the $\tilde{W}_{ij}$ given in Eq.~(\ref{css-W-sol}) with  
\begin{equation}
S(b,Q) = -\int_{c^2/b^2}^{Q^2}\, 
  \frac{d\bar{\mu}^2}{\bar{\mu}^2} \left[
  \ln\left(\frac{Q^2}{\bar{\mu}^2}\right) 
     A(\alpha_s(\bar{\mu})) + B(\alpha_s(\bar{\mu})) \right]\, .
\label{css-S}
\end{equation}
In Eq.~(\ref{css-W-sol}), all large logarithms from $\ln(c^2/b^2)$ to
$\ln(Q^2)$ in $\tilde{W}_{ij}(b,Q,x_A,x_B)$ are completely resummed
into the exponential factor $\exp[S(b,Q)]$, leaving the 
$\tilde{W}_{ij}(b,\frac{c}{b},x_A,x_B)$ with only one momentum scale
$1/b$. The $\tilde{W}_{ij}(b,\frac{c}{b},x_A,x_B)$ in
Eq.~(\ref{css-W-sol}) is then perturbatively calculable when 
the momentum scale $1/b$ is large enough, and is given by
\cite{CSS-W,Ellis-1}  
\begin{equation}
\tilde{W}_{ij}(b,\frac{c}{b},x_A,x_B) =
f_{i/A}(x_A,\mu=\frac{c}{b})\, f_{j/B}(x_B,\mu=\frac{c}{b})\, .
\label{css-W-pert}
\end{equation}
The functions $f_{i/A}$ and $f_{j/B}$ are the modified parton
distributions \cite{CSS-W,Ellis-1}, 
\begin{equation}
f_{i/A}(x_A,\mu) = \sum_a 
  \int_{x_A}^1\frac{d\xi}{\xi}\, 
  C_{i/a}(\frac{x_A}{\xi},\mu))\, \phi_{a/A}(\xi,\mu)
\label{mod-pdf}
\end{equation}
where $\sum_a$ runs over all parton flavors.  In Eq.~(\ref{mod-pdf}),
$\phi_{a/A}(\xi,\mu)$ is the normal parton distribution for finding a
parton of flavor $a$ in hadron $A$, 
and $C_{i/a}=\sum_{n=0} C_{i/a}^{(n)} (\alpha_s/\pi)^n$
are perturbatively calculable coefficient functions for finding a
parton $i$ from a parton $a$.  The first two coefficients of the
$C_{i/a}$ are available \cite{CSS-W,Davis}
\begin{eqnarray}
C^{(0)}_{i/j}(z,\mu=c/b) &=& \delta_{ij}\, \delta(z-1)\, ,
\nonumber\\
C^{(0)}_{i/g}(z,\mu=c/b) &=& 0\, ,
\nonumber \\
C^{(1)}_{i/j}(z,\mu=c/b) &=& \delta_{ij} \frac{C_F}{2} \left[ 
  (1-z) 
+ \left( \frac{\pi^2}{2}-4 \right) \delta(1-z) \right] ,
\nonumber \\
C^{(1)}_{i/g}(z,\mu=c/b) &=& T_R\, z\, (1-z)
\label{css-coef}
\end{eqnarray}
where $i$ and $j$ represent quark or antiquark flavors and $g$
represents a gluon.  The coefficient functions given in
Eq.~(\ref{css-coef}) are derived from the general functional forms in 
Ref.~\cite{CSS-W} by setting the renormalization constants and the
factorization scale as, $C_1=c$, $C_2=1$, and $\mu=c/b$.

The $\sigma_{ij\rightarrow V}(Q)$ in Eq.~(\ref{css-W-ij}) is the
lowest order cross section for a pair of quark and antiquark to
annihilate into the vector boson ($V=\gamma^*$, $W^{\pm},$ or $Z$).
For $V=\gamma^*$, we have
\begin{equation}
\sigma_{ij\rightarrow \gamma^*}(Q) = 
\delta_{j\bar{i}}\, e_i^2 
\left(\frac{4\pi^2\alpha_{EM}^2}{3 S}\right) 
\frac{1}{N}\, \frac{1}{Q^2}
\label{css-W-lo}
\end{equation}
where $e_i$ is the quark fractional charge and $N=3$ for SU(3) color.
The $\sigma_{ij\rightarrow V}(Q)$ for $V=W^{\pm}$ or $Z$ can be found
in Refs.~\cite{CSS-W,Ellis-1}.  

In the CSS resummation formalism, the $Y$ term in Eq.~(\ref{css-gen})
represents a small correction to the $Q_T$ distribution when $Q_T \ll
Q$.  But, it dominates the $Q_T$-distributions when $Q_T \sim Q$.  The
$Y$-term has a perturbative
expansion, $Y=\sum_{n=1} Y^{(n)} (\alpha_s(\mu)/\pi)^n$, and the
coefficients $Y^{(n)}$ have the following factorized form \cite{CSS-W} 
\begin{eqnarray}
Y^{(n)}(Q_T,Q,x_A,x_B;\mu) &=& \sum_{a,b} 
\int_{x_A}^1 \frac{d\xi_A}{\xi_A}\, \phi_{a/A}(\xi_A,\mu) 
\int_{x_B}^1 \frac{d\xi_B}{\xi_B}\, \phi_{b/B}(\xi_B,\mu) 
\nonumber\\
&\times &
\left(\frac{4\pi^2\alpha_{EM}^2}{9Q^2S}\right)
R_{ab\rightarrow V}^{(n)}(Q_T,Q,x_A/\xi_A,x_B/\xi_B;\mu)\, ,
\label{css-W-yn}
\end{eqnarray}
where $\sum_{a,b}$ run over all possible parton flavors and $\mu$
represents both the factorization and renormalization scale.
The $R_{ab\rightarrow V}^{(n)}$ in Eq.~(\ref{css-W-yn}) 
are perturbatively calculable and have the same normalization as those
introduced in Ref.~\cite{CSS-W}.  The leading order terms  
$R_{ab\rightarrow\gamma^*}^{(1)}$ are available and are given by
Eqs.~(2.9) to (2.12) in Ref.~\cite{CSS-W}.  For $W^{\pm}$ and $Z$
production, one needs to change the fractional quark charge $e_i^2$ in
the $R_{ab\rightarrow\gamma^*}^{(1)}$ by corresponding weak coupling 
constants \cite{CSS-W}.

Since the $Y$ term does not have the large logarithms and
perturbatively calculable, the predictive power of the CSS formalism
relies on our ability to predict the $\tilde{W}$ term in
Eq.~(\ref{css-gen}).  Because the leading power perturbative QCD
calculations and the normal parton distributions
in Eq.~(\ref{mod-pdf}) are only valid for $\mu > \mu_0 \sim 1-2$~GeV, 
the perturbatively calculated $b$-space distribution
$\tilde{W}(b,Q,x_A,x_B)$ in Eq.~(\ref{css-gen}) is reliable only if
the momentum scale $1/b > \mu_0$.  On the other hand, the Fourier
transform in Eq.~(\ref{css-gen}) requires a $b$-space distribution 
$\tilde{W}(b,Q,x_A,x_B)$ for $b\in [0,\infty)$.  Therefore, the
predictive power of the $b$-space resummation formalism is limited by
our inability to calculate the nonperturbative $b$-space distribution
at large $b$ \cite{QCD-rpt,Ellis-2}.

However, the $b$-space resummation formalism has a remarkable feature
that the resummed exponential factor $\exp[S(b,Q)]$ suppresses the
$b$-integral when $b$ is larger than $1/Q$.  Therefore, when $Q\gg
\mu_0$, it is possible that the Fourier transform in
Eq.~(\ref{css-gen}) is dominated by a region of $b$ much smaller than
$1/\mu_0$, and the calculated $Q_T$ distributions are insensitive to
the nonperturbative information at large $b$.  In fact, using the
saddle point method, it was shown \cite{PP-b,CSS-W} that for a large
enough $Q$, the QCD perturbation theory is valid even at $Q_T=0$, and
the Fourier transform in Eq.~(\ref{css-gen}) is dominated by an
impact parameter of order 
\begin{equation}
b_{\rm SP} = \frac{1}{\Lambda_{\rm QCD}}
  \left( \frac{\Lambda_{\rm QCD}}{Q} \right)^{\lambda}
\label{css-bsp}
\end{equation}
where $\lambda=16/(49-2n_f)\approx 0.41$ for quark flavors $n_f=5$.
From Eq.~(\ref{css-bsp}), the momentum scale corresponding to the
saddle point, $1/b_{\rm SP}$, can be well within the perturbative
region if the value of $Q$ is large enough.  Therefore, the predictive
power of the $b$-space resummation formalism is directly related to
the numerical value of the vector boson's invariant mass $Q$
\cite{PP-b,CSS-W}. 

For $W^{\pm}$ and $Z$ production, we have $Q\sim M_W$ or $M_Z$ and the  
corresponding momentum scale from Eq.~(\ref{css-bsp}), $1/b_{\rm SP}
\approx 10 \Lambda_{\rm QCD} \sim 2$~GeV, which is at the borderline
of the predictive power of perturbative QCD calculations without
introducing the power corrections.  In the rest of this section, we
show that the next-to-leading order corrections to the function
$S(b,Q)$ reduces the numerical value of the $b_{\rm SP}$.
Furthermore, we show that the numerical value for the saddle point has 
a strong dependence on the collision energy $\sqrt{S}$, and the
$\sqrt{S}$ dependence can either improves or reduces the predictive
power of the $b$-space resummation formalism.

Since there is no preferred transverse direction, the $\tilde{W}$ in
Eq.~(\ref{css-gen}) is a function of $b=|\vec{b}|$, and the Fourier
transform can be written as
\begin{eqnarray}
&\ &
\frac{1}{(2\pi)^2}\int d^2b\, e^{i\vec{Q}_T\cdot \vec{b}}\,
\tilde{W}(b,Q,x_A,x_B) 
\nonumber \\
&=& 
\frac{1}{2\pi} \int_0^\infty db\, b\, J_0(Q_T b)\, 
{\rm e}^{S(b,Q)}\,
\sum_{ij}\sigma_{ij\rightarrow V}(Q)\, 
\tilde{W}_{ij}(b,\frac{c}{b},x_A,x_B)\, ,
\label{css-W-F}
\end{eqnarray}
where $J_0(z)$ with $z=Q_T b$ is the Bessel function.  In deriving
Eq.~(\ref{css-W-F}), we used Eqs.~(\ref{css-W-ij}) and
(\ref{css-W-sol}).  The $b_{\rm SP}$ in Eq.~(\ref{css-bsp}) was
derived by solving 
\begin{equation}
\frac{d}{db}\ln\left(b\,{\rm e}^{S(b,Q)}\right)_{b=b_{\rm SP}} = 0
\label{css-saddle}
\end{equation}
with only $A^{(1)}$ for the function $S(b,Q)$.  Solving
Eq.~(\ref{css-saddle}) for the saddle point relies on the assumption
that the $b$-dependence in $\tilde{W}_{ij}(b,c/b,x_A,x_B)$ is smooth
around $b_{\rm SP}$.  

However, we find from Eq.~(\ref{css-W-pert}) that the $b$-dependence
in $\tilde{W}_{ij}(b,c/b,x_A,x_B)$ is strongly connected to the
numerical values of $x_A$ and $x_B$, and can be very important for
determining the saddle point if $x_A$ and $x_B$ are very small or very
large \cite{QZ1}.  Taking into account the full $b$-dependence of 
$\tilde{W}_{ij}(b,c/b,x_A,x_B)$, the saddle point for the
$b$-integration in Eq.~(\ref{css-W-F}) at $Q_T=0$ is determined by
solving the following equation, 
\begin{equation}
 \frac{d}{db}\ln\left(b\, {\rm e}^{S(b,Q)}\right)_{b=b_0}
+\frac{d}{db}\ln\left(\sum_{ij}\sigma_{ij\rightarrow V}(Q)\,
\tilde{W}_{ij}(b,\frac{c}{b},x_A,x_B)\right)_{b=b_0}
=0 .
\label{saddle}
\end{equation}
If the $\tilde{W}_{ij}(b,c/b,x_A,x_B)$ has a weak $b$-dependence
around $b_0$, the second term in Eq.~(\ref{saddle}) can be neglected,
and the $b_0\approx b_{\rm SP}$.  From Eq.~(\ref{css-W-pert}), the
$b$-dependence of $\tilde{W}_{ij}(b,c/b,x_A,x_B)$ is 
directly proportional to the evolution of the modified parton
distributions: 
\begin{eqnarray}
&\ & \frac{d}{db}\ln\left(\sum_{ij}\sigma_{ij\rightarrow V}(Q)\,
\tilde{W}_{ij}(b,\frac{c}{b},x_A,x_B)\right)
\nonumber \\
&\propto & 
- \frac{1}{b}\left[ 
\frac{d}{d\ln\mu} f_{i/A}(x_A,\mu) \quad
\mbox{or} \quad
\frac{d}{d\ln\mu} f_{j/B}(x_B,\mu) \right]\, ,
\label{mu-evo}
\end{eqnarray}
where $\mu = c/b$.  Since the coefficient functions $C$ in
Eq.~(\ref{css-coef}) do not have $b$-dependence at these orders, the  
evolution of the modified parton distributions in 
Eq.~(\ref{mu-evo}) is directly proportional to the evolution of normal
parton distributions, $(d/d\ln\mu)\,\phi_{i/A}(\xi,\mu)$.
Because of the steep falling feature of the normal parton
distributions when $\xi$ increases, the convolution over $\xi$ in
Eq.~(\ref{mod-pdf}) is dominated by the value of $\xi\sim x_A$.
Therefore, the evolution of the modified parton distributions in 
Eq.~(\ref{mu-evo}) is directly proportional to the evolution of normal
parton distributions, $(d/d\ln\mu)\,\phi_{i/A}(\xi,\mu)$ at
$\xi\sim x_A$.  From the DGLAP equation, it is known that
$(d/d\ln\mu)\phi(x,\mu)$ is positive (or negative) for $x<x_0\sim 0.1$
(or $x>x_0$), and the evolution is very steep when $x$ is far away
from $x_0$.  Therefore, the second term in Eq.~(\ref{saddle}) should
be very important when $x_A$ and $x_B$ are much smaller than the
$x_0$.  

\begin{figure}
\begin{center}
\epsfig{figure=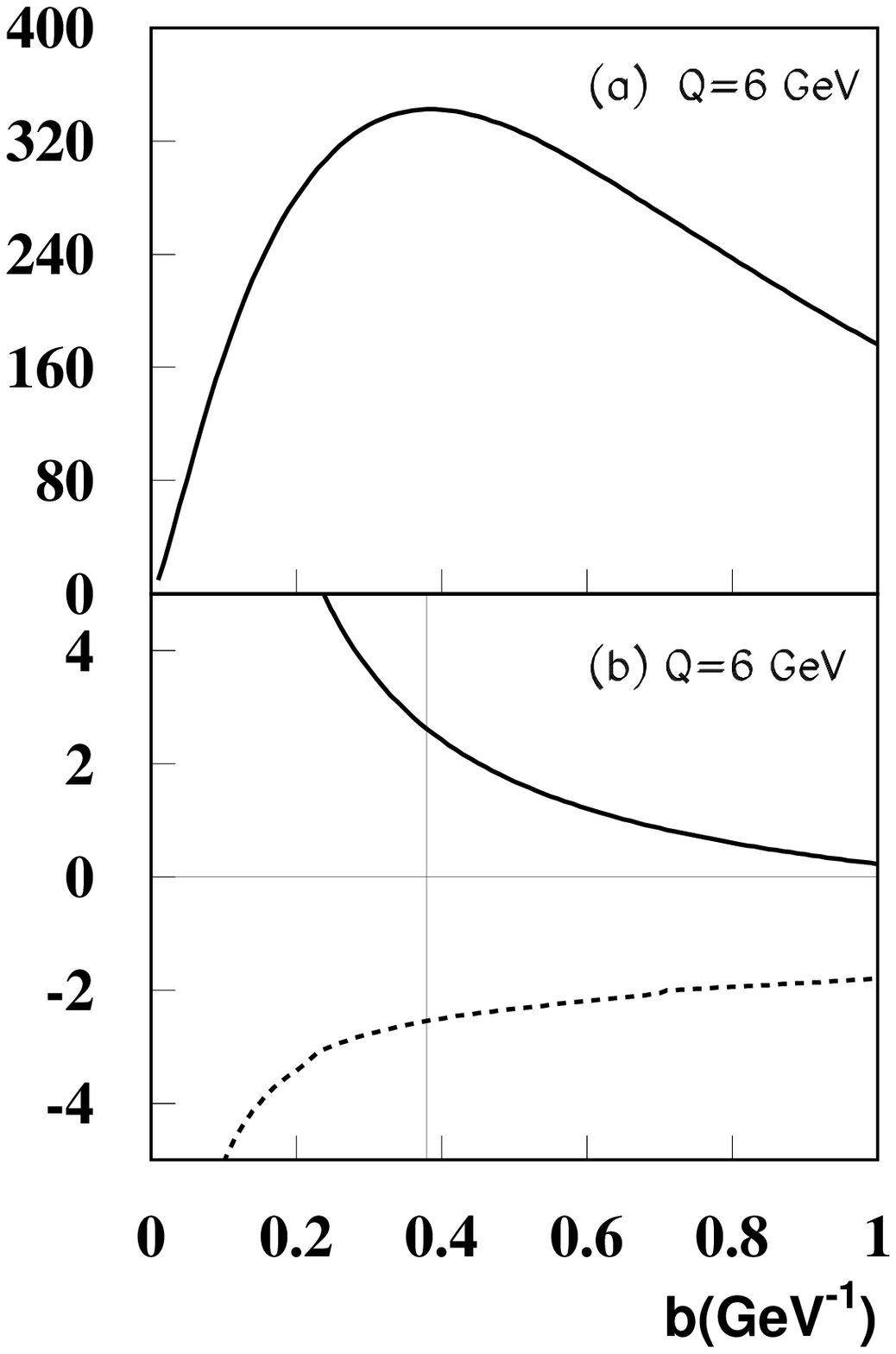,width=4.5in}
\end{center}
\caption{(a) Integrand of the $b$-integration in
Eq.~(\protect\ref{css-W-F}) at $Q_T=0$ and $Q=6$~GeV as a function of
$b$ with an arbitrary normalization at Tevatron energy;  
(b) the first (solid) and second
(dashed) terms in Eq.~(\protect\ref{saddle}) as a function of $b$ at
the same $Q$ and $\protect\sqrt{S}$.}
\label{fig1}
\end{figure}

Since Eq.~(\ref{css-saddle}) has a saddle point solution, the first
term in Eq.~(\ref{saddle}) is a decreasing function of $b$, and it
vanishes at $b=b_{\rm SP}$.  Because of the minus sign in
Eq.~(\ref{mu-evo}) and the fact that the number of small $x$ partons
increases when the scale $\mu$ increases, we expect the second
term in Eq.~(\ref{saddle}) to be negative when $x_A$ and $x_B$ are
smaller than the typical $x_0$, and to reduce the numerical value of
the saddle point.  As a demonstration, let $Q=6$~GeV and
$\sqrt{S}=1.8$~TeV.  Using CTEQ4M parton distributions and 
$\Lambda_{\rm QCD}(n_f=5)=0.202$~GeV \cite{CTEQ4}, one derives from
Eq.~(\ref{css-bsp}) that $b_{\rm SP}\approx 1.2$~GeV$^{-1}$, and might 
conclude that the perturbatively resummed $Q_T$ 
distribution at the given values of $Q$ and $\sqrt{S}$ is not
reliable.  However, as shown in Fig.~\ref{fig1}(a), the integrand of
the $b$-integration in Eq.~(\ref{css-W-F}) has a nice saddle point at
$b_0\approx 0.38$~GeV$^{-1}$, which is within the perturbative region.
This is due to the fact that $x_A\sim x_B\sim 0.003$ are very small.
The second term in Eq.~(\ref{saddle}) is negative and it
reduces the numerical value of the saddle point, which is clearly
shown in Fig.~\ref{fig1}(b).  The solid line and dashed line
represent the first and the second term in Eq.~(\ref{saddle}),
respectively.  Although the solid line in Fig.~\ref{fig1}(b) never
crosses zero for $b<1$~GeV$^{-1}$, which is consistent with the
fact that $b_{\rm SP}\sim 1.2$~GeV$^{-1}$, the dashed line is negative
and it cancels the solid line to give a nice saddle point at
$b=b_0\sim 0.38$~GeV$^{-1}$.   

\begin{figure}
\begin{center}
\epsfig{figure=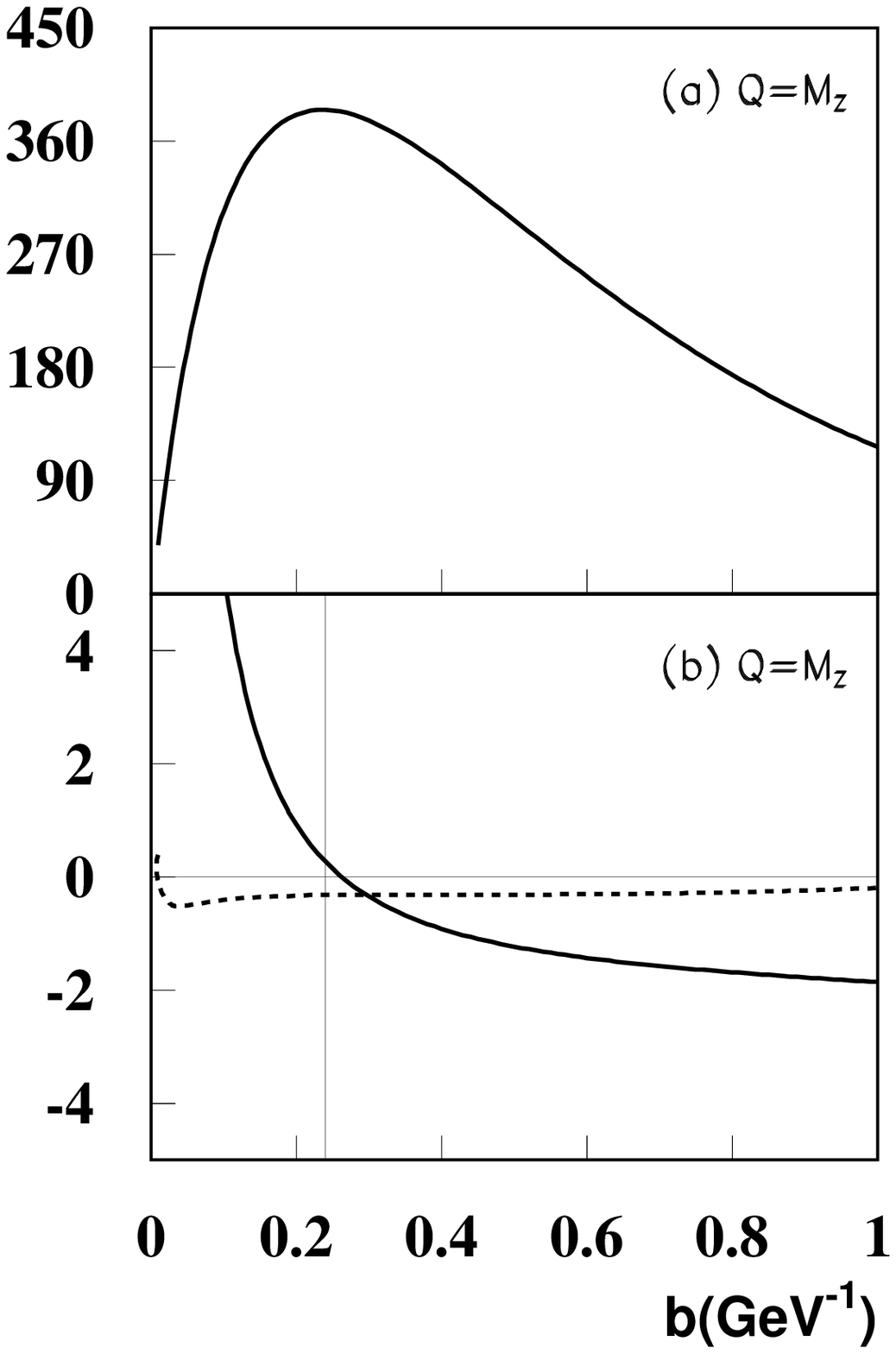,width=4.5in}
\end{center}
\caption{(a) Integrand of the $b$-integration in
Eq.~(\protect\ref{css-W-F}) at $Q_T=0$ and $Q=M_Z$ as a function of
$b$ with an arbitrary normalization at Tevatron energy
($\protect\sqrt{S}=1.8$~TeV); (b) the first
(solid) and second (dashed) terms in Eq.~(\protect\ref{saddle}) as a
function of $b$ at the same $Q$ and $\protect\sqrt{S}$.}
\label{fig2}
\end{figure}

\begin{figure}
\begin{center}
\epsfig{figure=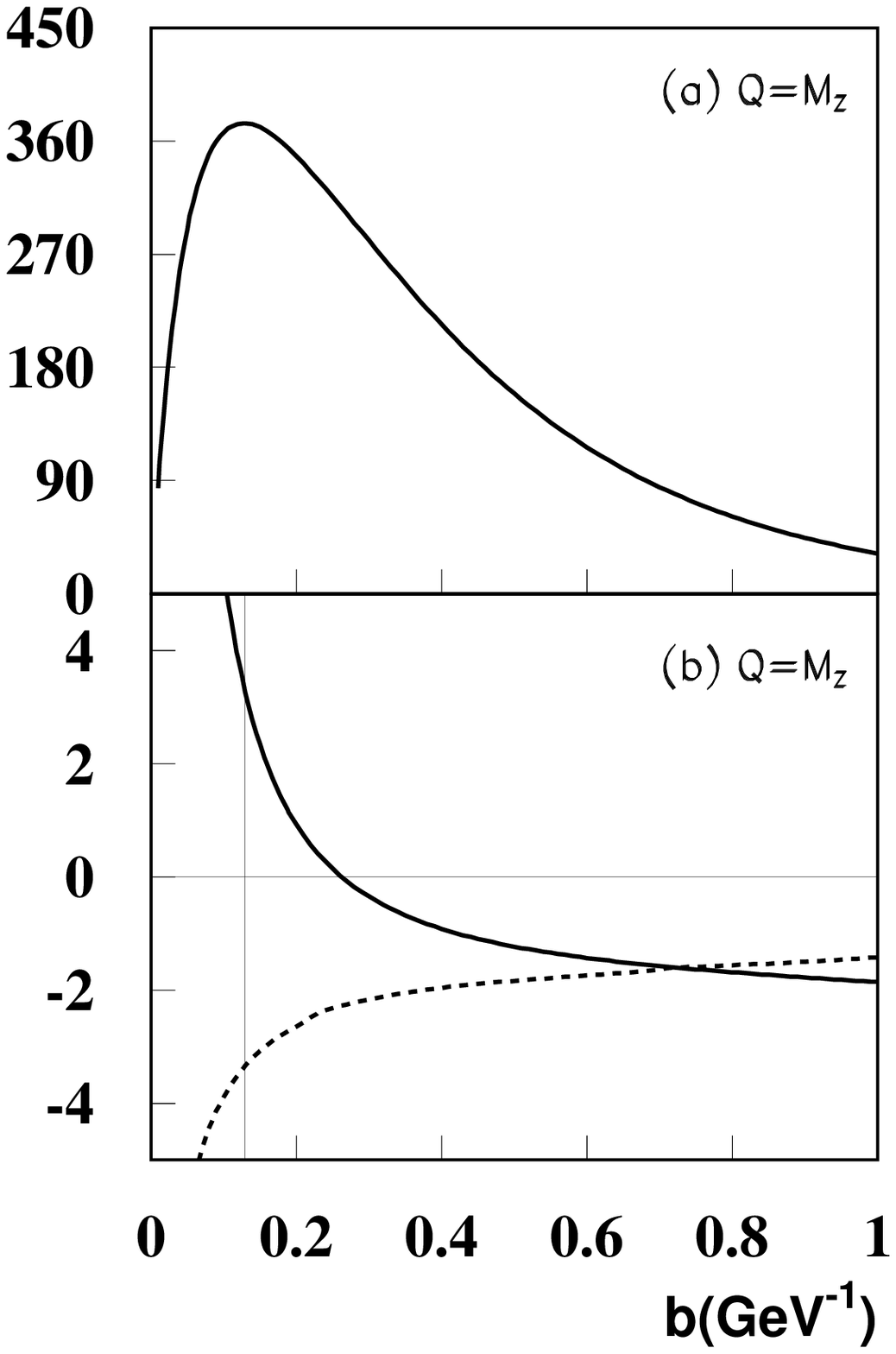,width=4.5in}
\end{center}
\caption{(a) Integrand of the $b$-integration in
Eq.~(\protect\ref{css-W-F}) at $Q_T=0$ and $Q=M_Z$ as a function of
$b$ with an arbitrary normalization at the LHC energy
($\protect\sqrt{S}=14$~TeV);    
(b) the first (solid) and second (dashed) terms in
Eq.~(\protect\ref{saddle}) as a function of $b$ at the same 
$Q$ and $\protect\sqrt{S}$.}
\label{fig3}
\end{figure}

Similar to Fig.~\ref{fig1}, we plot the integrand of the
$b$-integration in Eq.~(\ref{css-W-F}) for $Z$ production at Tevatron
and the LHC energies in Figs.~\ref{fig2} and \ref{fig3},
respectively.  In plotting both figures, we used CTEQ4M parton
distributions and the 
perturbatively calculated functions: $A$, $B$, and $C$ to the
next-to-leading order, which are listed in Eqs.~(\ref{css-AB}) and
(\ref{css-coef}). From Eq.~(\ref{css-bsp}), we estimate $b_{\rm
SP}\approx 0.4$~GeV$^{-1}$ for $Q=M_Z$.  As shown in
Fig.~\ref{fig2}(b), the solid  
line vanishes at $b\approx 0.27$~GeV$^{-1}$, which indicates that the
inclusion of the $A^{(2)}$, $B^{(1)}$, and $B^{(2)}$ reduces the
numerical value of the saddle point, $b_{\rm SP}$.  The dashed line in
Fig.~\ref{fig2}(b), which corresponds to the second term in
Eq.~(\ref{saddle}), further reduces the numerical value of the saddle
point to $b_0\approx 0.24$~GeV$^{-1}$.  At the LHC energy, $x_A$ and
$x_B$ are much smaller.  We then expect the second term in 
Eq.~(\ref{saddle}) to be more important, which is clearly shown
in Fig.~\ref{fig3}(b).  The dashed line in
Fig.~\ref{fig3}(b) has a much larger absolute value in comparison
with that in Fig.~\ref{fig2}(b).  Consequently, the numerical value of
the saddle point is further reduced from $b_0\approx 0.24$~GeV$^{-1}$
at $\sqrt{S}=1.8$~TeV to $b_0\approx 0.13$~GeV$^{-1}$ at 
$\sqrt{S}=14$~TeV, where the perturbative QCD calculations should be
reliable.  In addition, the narrow width of the $b$-distribution shown
in Fig.~\ref{fig3}(a) ensures that the $b$-integration is dominated by
$b\sim b_0$.  In conclusion, even at $Q_T=0$, the perturbative QCD
based $b$-space resummation formalism is valid as long as the
collision energy $\sqrt{S}$ is large enough.

When $Q_T>0$, the Bessel function $J_0(z=Q_T b)$ further suppresses
the large $b$ region of the $b$-integration.  Because the argument of
the Bessel function is proportional to $Q_T$, the large $b$ region is
more suppressed if $Q_T$ is larger.  That is, the larger $Q_T$ is, the
better the $b$-space resummation formalism is expected to work.
However, it has been known \cite{QCD-rpt} that the $b$-space resummed
$Q_T$ distribution from Eq.~(\ref{css-gen}) becomes unphysical or
even negative when $Q_T$ is large.  For example, a matching between
the resummed and fixed-order calculations has to take place at $Q_T
\sim 50$~GeV for $W^{\pm}$ production when these two predictions cross
over \cite{Ellis-2}.  We will address this puzzle in Sec.~\ref{sec4}.

\begin{figure}
\begin{center}
\epsfig{figure=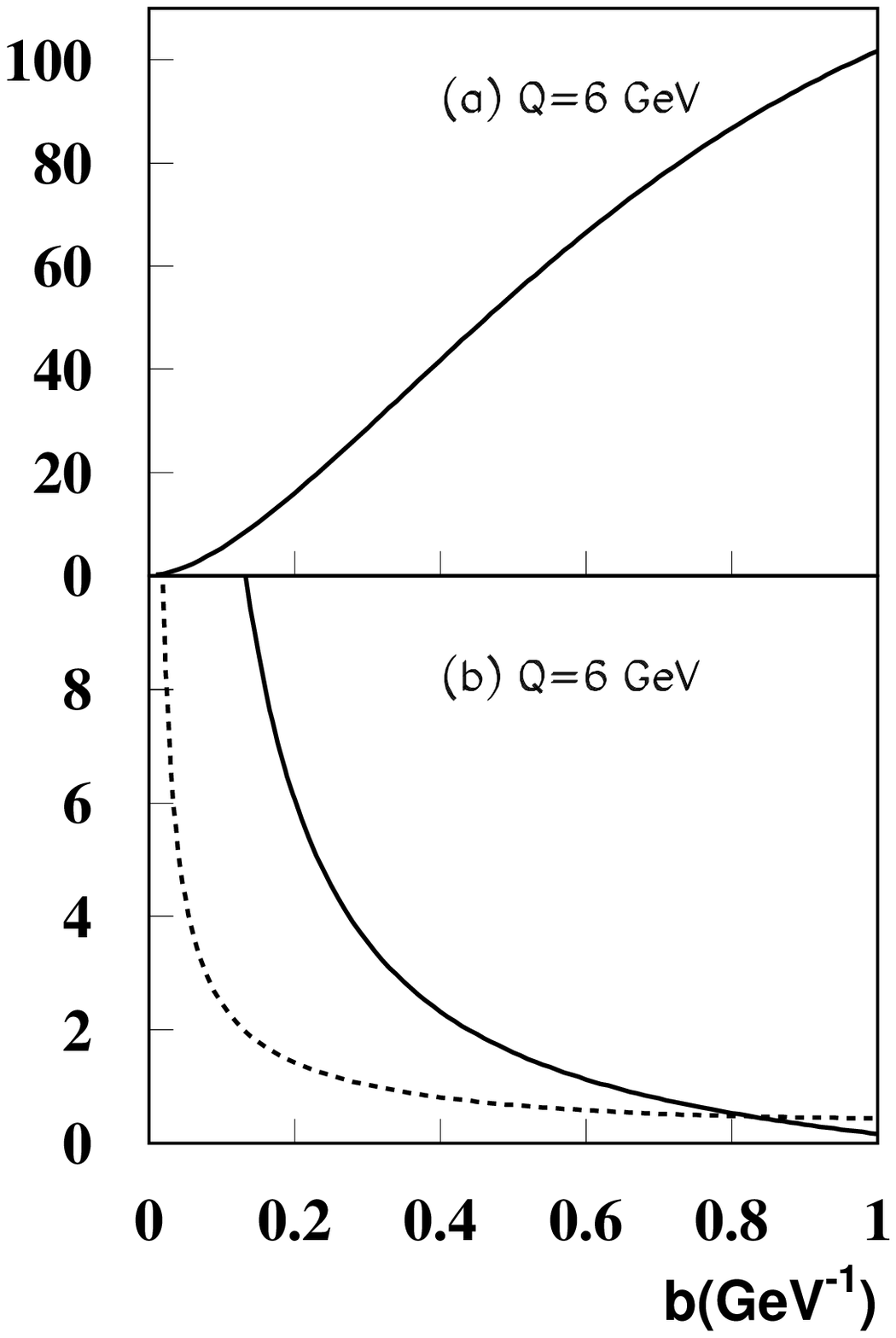,width=4.5in}
\end{center}
\caption{(a) Integrand of the $b$-integration in
Eq.~(\protect\ref{css-W-F}) at $Q_T=0$ and $Q=6$~GeV as a function of
$b$ with an arbitrary normalization at E288 energy 
($\protect\sqrt{S}=27.4$~GeV);   
(b) the first (solid) and second (dashed) terms in
Eq.~(\protect\ref{saddle}) as a function of $b$ at the same 
$Q$ and $\protect\sqrt{S}$.}
\label{fig4}
\end{figure}

In the rest of this section, we investigate the predictive power of
the $b$-space resummation formalism for the Drell-Yan production at
fixed target energies ($\sqrt{S}\leq 40$~GeV).  Most data at the fixed
target energies have $Q \in (5, 12)$~GeV and $Q_T$ at a few GeV or
less.  From Eq.~(\ref{css-bsp}), we find that $b_{\rm SP}$ is of order
1~GeV$^{-1}$ or larger.  Because of the low collision energy, the
typical values of $x_A$ and $x_B$ are larger than the $x_0$. 
Therefore, the second term in the Eq.~(\ref{saddle}) should be
positive, which increases the numerical value of the saddle point.  As
an illustration, instead of $\sqrt{S}=1.8$~TeV, we replot all
quantities in Fig.~\ref{fig1} at $\sqrt{S}=27.4$~GeV in
Fig.~\ref{fig4}, which is the collision energy for Fermilab experiment
E288 \cite{DY-E288}.  
As expected, the dashed line is now positive and the saddle point
is no longer in the perturbative region.  In conclusion, at fixed
target energies, the perturbatively calculated $b$-space distribution
derived from the CSS resummation formalism is not sufficient to
predict the Drell-Yan $Q_T$ distributions at $Q_T=0$.  A
nonperturbative extrapolation to the large $b$ region is necessary.  

\begin{figure}
\epsfig{figure=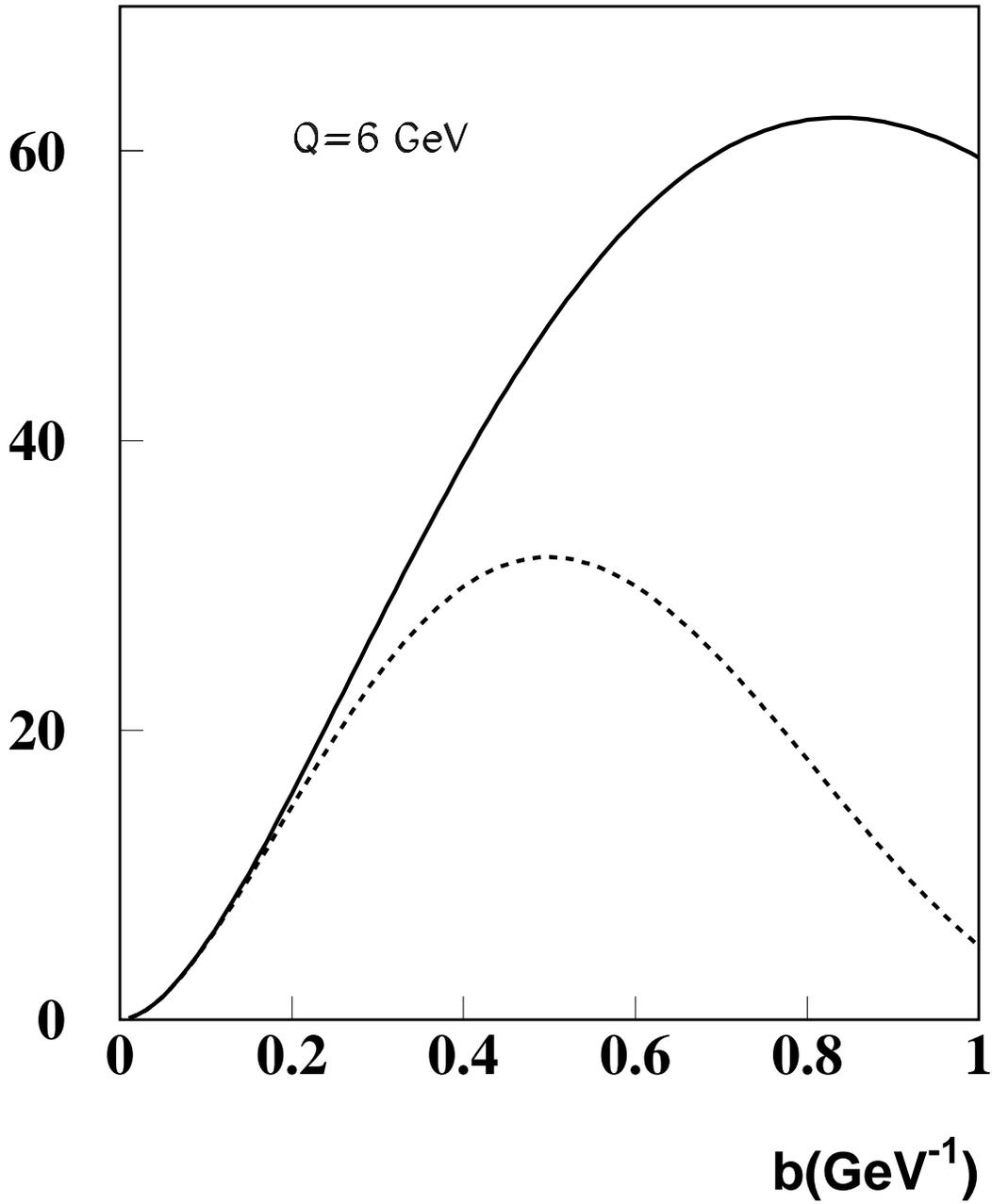,width=5.5in}
\caption{Integrand of the $b$-integration in
Eq.~(\protect\ref{css-W-F}) at $Q_T=1$~GeV (solid line) and 
$Q_T=2$~GeV (dashed line) as a function of $b$ with an
arbitrary normalization.  The $Q$ and $\protect\sqrt{S}$ are the same
as those in Fig.~\protect\ref{fig4}(a).}
\label{fig5}
\end{figure}

When $Q_T>0$, the Bessel function $J_0(z=Q_T b)$ suppresses
the large $b$ region of the $b$-integration, and improves the
predictive power of the $b$-space resummation formalism at the fixed
target energies.  In Fig.~\ref{fig5}, we plot the integrand of the
$b$-integration in Eq.~(\ref{css-W-F}) at $Q_T=1$~GeV
(solid line) and $Q_T=2$~GeV (dashed line) for $Q=6$~GeV and
$\sqrt{S}=27.4$~GeV.  As shown in
Fig.~\ref{fig5}, the saddle point for the $b$-integration moves to the
smaller $b$ region as $Q_T$ increases.  Since the saddle points of
both curves in Fig.~\ref{fig5} are within the perturbative region, one
might expect the $b$-space resummation formalism to provide a good
description of the $Q_T$ distributions at these energies.  However,   
due to the oscillatory nature of the Bessel function, the precise
value of the $b$-integration depends on the detailed cancellations in
the large $b$ region.  Therefore, the predictive power of the
$b$-space resummation formalism at the fixed target energies is still
limited by our knowledge of the nonperturbative information at large
$b$.   More discussions are given in Sec.~\ref{sec4}.


\section{Extrapolation to the Large $b$ Region} 
\label{sec3}

In this section, we quantitatively analyze the role of the
nonperturbative input at large $b$ in the $b$-space resummation
formalism.  We first briefly review the extrapolation defined in
Eq.~(\ref{css-W-b}) and its status in comparison with the existing
data.  Then, by adding possible power corrections to the
renormalization group equations in Eqs.~(\ref{css-K-rg}) and
(\ref{css-G-rg}), we derive a new functional form for extrapolating
the perturbatively resummed $\tilde{W}(b,Q,x_A,x_B)$ to the large $b$ 
region.  This new functional form clearly separates the
perturbative prediction at small $b$ from the nonperturbative
physics in the large $b$ region.  

\subsection{Extrapolation proposed by Collins, Soper, and Sterman} 
\label{sec3a}

As discussed in last section, the perturbatively resummed
$\tilde{W}(b,Q,x_A,x_B)$ in Eq.~(\ref{css-W-ij}) is only reliable for
the small $b$ region.  An extrapolation of the perturbatively
calculated $\tilde{W}(b,Q,x_A,x_B)$ to the large $b$ region is
necessary in order to complete the Fourier transform in
Eq.~(\ref{css-gen}).  In Ref.~\cite{CSS-W}, CSS proposed the following 
extrapolation,
\begin{equation}
\tilde{W}^{\rm CSS}(b,Q,x_A,x_B) \equiv \sum_{ij}
\sigma_{ij\rightarrow V}(Q) \,
\tilde{W}_{ij}(b_*,Q,x_A,x_B)\,
F^{NP}_{ij}(b,Q,x_A,x_B)\, ,
\label{css-fnp-ij}
\end{equation}
where $b_*$ was defined following Eq.~(\ref{css-W-b}), and the 
perturbatively calculated $\tilde{W}_{ij}(b,Q,x_A,x_B)$ are given in
Eq.~(\ref{css-W-sol}).  The nonperturbative input distributions
$F_{ij}^{NP}$ have the following functional form \cite{CSS-W}
\begin{equation}
F^{NP}_{ij}(b,Q,x_A,x_B) = \exp\left[
-\ln(Q^2/Q_0^2)\, g_1(b) - g_{i/A}(x_A,b) - g_{j/B}(x_B,b) \right]
\label{css-fnp}
\end{equation}
where the $\ln(Q^2)$ dependence is a derived result.  The functions
$g_1(b)$, $g_{i/A}(x_A,b)$, and $g_{j/B}(x_B,b)$ are nonperturbative,
and should go to zero as $b\rightarrow 0$.  The predictive power of
the CSS formalism relies on the derived $Q^2$ dependence 
and the universality of 
the $F_{ij}^{NP}$.  Since the low energy Drell-Yan data are sensitive
to the large $b$ region, in principle, one can use the low $Q^2$ data
to fix the parameters of the nonperturbative $F^{NP}_{ij}$ and predict
the $Q_T$ distributions of $W^{\pm}$ and $Z$ production at high $Q^2$.

Davis, Webber, and Stirling (DWS) introduced the following form for
the nonperturbative distribution $F^{NP}_{ij}$ \cite{Davis} 
\begin{equation}
F^{NP}_{DWS}(b,Q,x_A,x_B) 
= \exp\left[-b^2\left(g_1 + g_2 \ln(Q/2Q_0)\right)\right] \, ,
\label{dws-fnp}
\end{equation}
where $Q_0=2$~GeV~$=1/b_{max}$, and $g_1$ and $g_2$ are flavor
independent fitting parameters.  Without the flavor dependence, the
extrapolated CSS formalism defined in Eq.~(\ref{css-fnp-ij}) reduces
to that in Eq.~(\ref{css-W-b}).  With $g_1=0.15$~GeV$^{2}$ and
$g_2=0.4$~GeV$^{2}$, DWS found \cite{Davis} that the CSS $b$-space
resummation formalism gives a reasonable description of the Drell-Yan
data from Fermilab experiment E288 at $\sqrt{S}=27.4$~GeV
\cite{DY-E288} as well as CERN ISR experiment R209 at
$\sqrt{S}=62$~GeV \cite{DY-R209}.   

In order to incorporate possible $\ln(\tau)$ dependence with
$\tau=Q^2/S=x_A x_B$, Ladinsky and Yuan (LY) proposed a modified
functional form for the $F^{NP}_{ij}$\cite{LY} 
\begin{equation}
F^{NP}_{LY}(b,Q,x_A,x_B) 
= \exp\left[-b^2\left(g_1 + g_2 \ln(Q/2Q_0)\right) 
- b\,g_1\,g_3\,\ln(100 x_A x_B)\right] \, .
\label{ly-fnp}
\end{equation}
An extra parameter $g_3$ was introduced in LY's parameterization of the
nonperturbative $F^{NP}_{ij}$.  Similar to DWS' parameterization, no
flavor dependence was introduced into the nonperturbative
distribution.  With $g_1=0.11^{+0.04}_{-0.03}$~GeV$^2$, 
$g_2=0.58^{+0.1}_{-0.2}$~GeV$^2$, and 
$g_3=-0.15^{+0.1}_{-0.1}$~GeV$^{-1}$, LY was able to fit the 
R209 Drell-Yan data as well as CDF data on $W$ and $Z$ production from
Fermilab.  More recently, Landry, Brock, Ladinsky, and Yuan (LBLY)
performed a much more extensive global fit to the low energy Drell-Yan
data as well as high energy $W$ and $Z$ data by using both DWS and LY
parameterizations \cite{LBLY}.  In order to fit both the low energy 
Drell-Yan and the collider $W$ and $Z$ data, LBLY found that it is
necessary to introduce a large overall normalization uncertainty in
order to include the low energy Drell-Yan data (in
particular, E288 data) into the global fit \cite{LBLY}.  LBLY also
emphasized that the collider data on $Z$ production are very useful in
determining the value of fitting parameter $g_2$.  They concluded
\cite{LBLY} that both DWS and LY parameterizations with updated
parameters result in good global fits, but give measurable differences
in $Q_T$ distributions of $Z$ production at Tevatron.

\begin{figure}
\begin{center}
\epsfig{figure=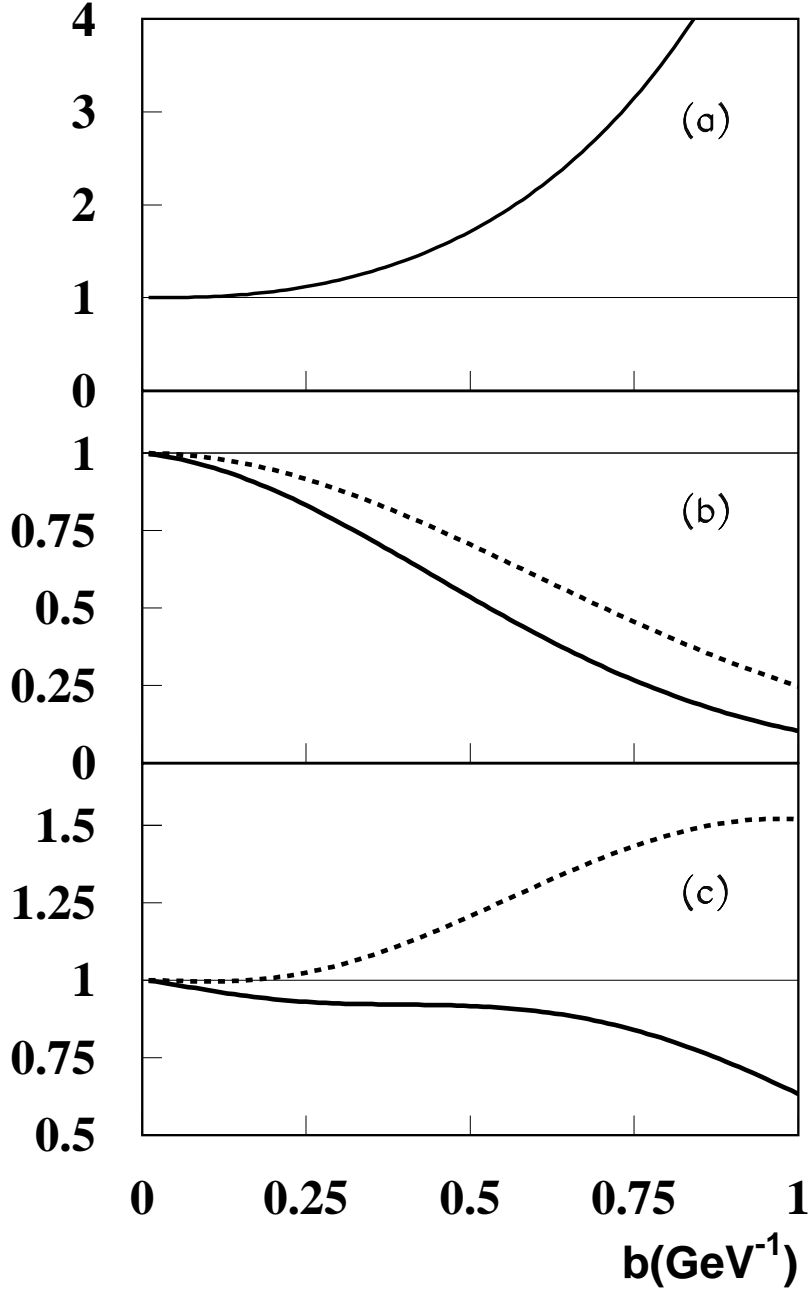,width=4.3in}
\end{center}
\caption{(a) Ratio $\tilde{W}(b_*,Q,x_A,x_B)/\tilde{W}(b,Q,x_A,x_B)$
as a function of $b$; (b) $F^{NP}(b,Q,x_A,x_B)$ as a function of $b$;
(c) Ratio $R_{W}$ defined in Eq.~(\protect\ref{rw}) as a function of
$b$.  All plots have $Q=M_Z$ and $\protect\sqrt{S}=1.8$~TeV.}
\label{fig6}
\end{figure}

Based on our discussions in last section, the $Q_T$ distributions of
$Z$ production at collider energies should not be very sensitive to
the nonperturbative physics from the large $b$ region.  Any significant
dependence on the fitting parameters for $Z$ production would cast a
doubt on the predictive power of the $b$-space resummation
formalism.  To understand the fitting parameter dependence of the $Z$
production, we introduce the following ratio,
\begin{eqnarray}
R_W(b,Q,x_A,x_B)
&\equiv & 
\frac{\tilde{W}^{\rm CSS}(b,Q,x_A,x_B)}
     {\tilde{W}(b,Q,x_A,x_B)}
\nonumber \\
&=&
\frac{\tilde{W}(b_*,Q,x_A,x_B)}
     {\tilde{W}(b,Q,x_A,x_B)} \, 
F^{NP}(b,Q,x_A,x_B)\, ,
\label{rw}
\end{eqnarray}
where $\tilde{W}(b,Q,x_A,x_B)$ is the perturbatively calculated
$b$-space distribution given in Eq.~(\ref{css-W-ij}).  In deriving the
second line in Eq.~(\ref{rw}), we used the fact that both DWS and LY
parameterizations of the $F^{NP}$ are independent of the parton
flavors.  Using CTEQ4M parton distributions, we plot in
Fig.~\ref{fig6}(a) the ratio 
$\tilde{W}(b_*,Q,x_A,x_B)/\tilde{W}(b,Q,x_A,x_B)$ as a function of $b$
at $Q=M_Z$ and $\sqrt{S}=1.8$~TeV.  In Fig.~\ref{fig6}(b), we plot the
nonperturbative distribution $F^{NP}(b,Q,x_A,x_B)$ as a function of
$b$ with both DWS parameters (dashed line) and LY parameters (solid
line).  In Fig.~\ref{fig6}(c), we plot the ratio $R_W(b,Q,x_A,x_B)$
defined in Eq.~(\ref{rw}), which is effectively equal to a product of
Fig.~\ref{fig6}(a) and Fig.~\ref{fig6}(b).  From Fig.~\ref{fig6}, we
learn that the introduction of the $b_*$ significantly changes the
perturbatively calculated $b$-space distribution within the
perturbative region; and in the same region, 
the function $F^{NP}$ can deviate from the unity by as
much as 50 percents for some fitting parameters.  In order to
preserve the predictive power of the perturbative calculations,
it is important to keep the $\tilde{W}^{\rm CSS}(b,Q,x_A,x_B)$
consistent with the perturbatively calculated
$\tilde{W}(b,Q,x_A,x_B)$ when $b<b_{max}$ (or 
$R_W(b,Q,x_A,x_B)\approx 1$).  However, we find that a
significant fitting parameter dependence (as much as 20 percents) was
introduced by the $\tilde{W}^{\rm CSS}(b,Q,x_A,x_B)$
to the $b$-space distribution within the 
perturbative region.  The same conclusion holds if we
plot the curves at different energies or use other sets of fitting
parameters available for the $F^{NP}$ \cite{Ellis-1,LBLY}.

\subsection{Extrapolation with dynamical power corrections}
\label{sec3b}

In order to separate the perturbative prediction in the small $b$
region from the nonperturbative physics at large $b$, we derive a new
functional form to extrapolate the perturbatively calculated 
$\tilde{W}(b,Q,x_A,x_B)$ to the large $b$ region.  Our goal is to have
an extrapolation that preserves the predictive power of perturbative
calculations in the small $b$ region and extends to the large $b$
region with as much correct physics as we can put in.  

Taking advantage of our early conclusion that heavy boson production
at collider energies should not be very sensitive to the large $b$
region, we can improve the leading power perturbative QCD calculations
by studying the behavior of power corrections in the region of
$b$-space where $b$ is not too much larger than $b_{max}$.  The power
correction in QCD is a very rich and difficult subject itself
\cite{PW-OPE,PW-DIS,PW-DY,PW-Renorm}.  
In order to define the power corrections, we have to
identify a nonperturbative momentum scale, $\Lambda$, which should be
of order $\Lambda_{\rm QCD}$.  For example, the nonperturbative scale
can be the target mass \cite{PW-mass} or matrix elements of high twist
operators \cite{PW-OPE,PW-DIS,PW-DY}.  In 
addition, we have to distinguish two different types of power
corrections: (1) the power corrections directly to the
physical observables (such as cross sections or structure functions),
and (2) the power corrections to the evolution or renormalization
group equations.  
The type-one (or direct) power corrections are always proportional to
the power of $(\Lambda/Q)$ with the physically observed momentum scale
$Q$ \cite{PW-OPE,PW-DIS,PW-DY}.  
Therefore, the effect of this type of power
corrections to physical observables can be neglected when
$(\Lambda/Q)\rightarrow 0$.  Similarly, the type-two (or indirect)
power corrections are proportional to the power of $(\Lambda/\mu)$
with evolution or renormalization scale $\mu$.  It is important to
note that physical 
observables are not directly proportional to the evolution or
renormalization group equations, instead, they depend on the {\it
solutions} of these equations.  Therefore, physical
observables carry the effect of the type-two power corrections for
all $\mu \in [Q_0,Q]$ and the boundary conditions at the scale $Q_0$
\cite{PW-evo-MQ}.  Even when $Q$ is much larger than $\Lambda$,
physical observables can still carry a large effect of the type-two
power corrections 
through the evolution from $Q_0$ to $Q$ \cite{PW-evo-Qiu}.  In this
subsection, we concentrate on the type-two power corrections.
Because they contribute to the evolution or renormalization
group equations, these power corrections should have a dynamical
origin.   

When $b>b_{max}$, we solve the evolution equation in
Eq.~(\ref{css-W-evo}) from $\ln(c^2/b_{max}^2)$ to $\ln(Q^2)$ in order
to separate the leading power QCD calculations at small $b$ from the
large $b$ region.  Because the scale $Q$ of the evolution equation is
chosen to be larger than $c/b_{max}$, we can ignore the explicit
$1/Q^2$ power corrections to this equation.  

However, the kernel
$K(b\mu,\alpha_s(\mu))$ of the evolution equation has an explicit
$b$-dependence, we need to add power corrections to its
renormalization group equation when $b>b_{max}$.  Similarly, we need
to add power corrections to the renormalization group equation of the
kernel $G(Q/\mu,\alpha_s(\mu))$ when $1/\mu > b_{max}$.  
Since we are only interested in deriving a functional form
of the $b$-dependence due to power corrections, we will not attempt to
derive the exact coefficients of the power corrections to
the renormalization group equations by going through detailed analysis
on the mixing of leading and high twist operators \cite{PW-evo-MQ}. 
Instead, we introduce some fitting parameters for the size of the
possible power corrections.  For including only the leading power
corrections, we modify the renormalization group equations 
in Eqs.~(\ref{css-K-rg}) and (\ref{css-G-rg}) as follows,
\begin{eqnarray}
\frac{\partial}{\partial \ln\mu^2} K(b\mu,\alpha_s(\mu))
&=& -\frac{1}{2}\, \gamma_K(\alpha_s(\mu)) 
    -\frac{1}{\mu^2}\, \bar{\gamma}_K\, ,
\label{qz-K-rg} \\
\frac{\partial}{\partial \ln\mu^2} G(Q/\mu,\alpha_s(\mu))
&=& \frac{1}{2}\, \gamma_K(\alpha_s(\mu)) 
    +\frac{1}{\mu^2}\, \bar{\gamma}_K\, ,
\label{qz-G-rg}
\end{eqnarray}
where $\bar{\gamma}_K$ is treated as an unknown
parameter here, though it should in principle depend on
$\alpha_s(\mu)$ and the characteristic size of high twist operators.
In Eqs.~(\ref{qz-K-rg}) and (\ref{qz-G-rg}), we parameterize the
leading power corrections in such a way that they preserve 
$(d/d\ln\mu^2) \tilde{W}_{ij}(b,Q,x_A,x_B)=0$.  Since we are
interested in the region of $b$ not too much larger than $b_{max}$, we
neglect higher power corrections in Eqs.~(\ref{qz-K-rg}) and
(\ref{qz-G-rg}). This approximation is going to be tested in
Sec.~\ref{sec4}. 

By integrating Eq.~(\ref{qz-K-rg}) over $\ln(\mu^2)$ from
$\ln(c^2/b^2)$ to $\ln(\mu^2)$ and Eq.~(\ref{qz-G-rg}) from $\ln(Q^2)$
to $\ln(\mu^2)$, we have 
\begin{eqnarray}
K(b\mu,\alpha_s(\mu))+G(Q/\mu,\alpha_s(\mu)) \approx 
&-&
\int_{c^2/b_{max}^2}^{Q^2}\, 
  \frac{d\bar{\mu}^2}{\bar{\mu}^2}\, 
  A(\alpha_s(\bar{\mu}))
- B(\alpha_s(Q))
\nonumber \\
&-&
\int_{c^2/b^2}^{c^2/b_{max}^2}\, 
  \frac{d\bar{\mu}^2}{\bar{\mu}^2}\, 
  \left[\frac{1}{2}\gamma_K(\alpha_s(\bar{\mu}))
       +\frac{1}{\bar{\mu}^2}\,\bar{\gamma}_K\right]
\nonumber \\
&+&\left[K(c,\alpha_s(\frac{c}{b}))
      -K(c,\alpha_s(\frac{c}{b_{max}}))\right]\, ,
\label{qz-KG}
\end{eqnarray}
where $A$ and $B$ are the same as those defined in
Eq.~(\ref{css-KG}).  In deriving Eq.~(\ref{qz-KG}), we neglected the
power corrections for the momentum scale between $c^2/b_{max}^2$ and
$Q^2$, which is consistent with neglecting the $1/Q^2$ term in the
evolution equation of the $\tilde{W}_{ij}(b,Q,x_A,x_B)$.  Substituting
the $K$ and $G$ in Eq.~(\ref{qz-KG}) into the evolution equation in 
Eq.~(\ref{css-W-evo}) and solving the evolution equation over the
$\ln(Q^2)$ from $\ln(c^2/b_{max}^2)$ to $\ln(Q^2)$, we obtain the
solution for $b>b_{max}$,
\begin{equation}
\tilde{W}^{QZ}_{ij}(b,Q,x_A,x_B)=
  \tilde{W}_{ij}(b_{max},Q,x_A,x_B)\,
  \tilde{F}^{NP}_{ij}(b,Q,x_A,x_B;b_{max})
\label{qz-W-ij}
\end{equation}
where $\tilde{W}_{ij}$ is the leading power perturbative solution
given in Eq.~(\ref{css-W-sol}), and 
\begin{eqnarray}
\tilde{F}^{NP}_{ij}(b,Q,x_A,x_B;b_{max}) 
&=&
\frac{\tilde{W}_{ij}(b,c/b_{max},x_A,x_B)}
      {\tilde{W}_{ij}(b_{max},c/b_{max},x_A,x_B)}
\nonumber \\
&\times &
\exp\Bigg\{ -\ln\left(\frac{Q^2 b_{max}^2}{c^2}\right) \left[
    \frac{\gamma}{\alpha}
         \left( (b^2)^\alpha - (b_{max}^2)^\alpha\right)
   +\frac{\bar{\gamma}_K}{c^2}\left( b^2 - b_{max}^2\right)
\right.
\nonumber \\
&\ & \left. {\hskip 1.4in}
   -\left(K(c,\alpha_s(\frac{c}{b}))
      -K(c,\alpha_s(\frac{c}{b_{max}}))\right) \right] \Bigg\}.
\label{qz-fnp-ij}
\end{eqnarray}
The nonperturbative function $\tilde{F}^{NP}_{ij}\rightarrow 1$ as
$b\rightarrow b_{max}$.  In deriving Eq.~(\ref{qz-fnp-ij}), we
approximate the $\mu$-dependence of $\gamma_K(\alpha_s(\mu))$ in the
small  
$\mu$ region by $\frac{1}{2}\gamma_K(\alpha_s(\mu)) \approx \gamma\,
(\mu^2)^{-\alpha}$ with constant parameters $\gamma$ and $\alpha$, and 
we expect $\alpha$ to be much less than one.  This approximation is to
mimic a summation of a perturbative series in powers of the running
coupling constant, $(\alpha_s(\mu))^m$, with the scale $\mu$
extrapolated into the nonperturbative region \cite{AMS-DY}.  Since the 
$[K(c,\alpha_s(\frac{c}{b}))-K(c,\alpha_s(\frac{c}{b_{max}}))]$
in Eq.~(\ref{qz-fnp-ij}) depends only on $b$ and $b_{max}$ through
$\alpha_s$, we can combine it with the first term and treat the power
$\alpha$ and the coefficient as fitting parameters.  Since
the $b$-dependence of the $\tilde{W}_{ij}(b,c/b_{max},x_A,x_B)$ 
depends on the evolution of parton distributions, the 
$\ln[\tilde{W}_{ij}(b,c/b_{max},x_A,x_B)/
\tilde{W}_{ij}(b_{max},c/b_{max},x_A,x_B)]$ changes sign when $x_A$
and $x_B$ are larger or smaller than the typical $x_0$.  Therefore,
we can parameterize the ratio
\begin{equation}
\frac{\tilde{W}_{ij}(b,c/b_{max},x_A,x_B)}
     {\tilde{W}_{ij}(b_{max},c/b_{max},x_A,x_B)}
\approx \exp\left\{
      g_3\ln\left(\frac{x_A x_B}{x_0^2}\right)
            \left((b^2)^\beta-(b_{max}^2)^\beta\right) \right\}
\label{qz-xab}
\end{equation}
with parameters $g_3$ and $\beta$.  In principle, the parameters
$g_3$ and $\beta$ as well as $x_0$ can depend on parton flavors
because of the flavor dependence of parton evolutions.  Since
parton distributions are near saturation at a very small momentum
scale \cite{AHM-Satu}, we expect both $g_3$ and $\beta$ to be very
small, and therefore, this term can be neglected in comparison with
other terms in the $\tilde{F}^{NP}_{ij}$.  Consequently, we can
neglect the flavor dependence of the $\tilde{F}^{NP}_{ij}$.  In
conclusion, without losing the characteristic features of the
$b$-dependence in Eq.~(\ref{qz-fnp-ij}), we can reparameterize the
$\tilde{F}^{NP}_{ij}$ as  
\begin{equation}
\tilde{F}^{NP}_{QZ}(b,Q,x_A,x_B;b_{max}) =
\exp\left\{ -\ln(\frac{Q^2 b_{max}^2}{c^2}) \left[
    g_1 \left( (b^2)^\alpha - (b_{max}^2)^\alpha\right)
   +g_2 \left( b^2 - b_{max}^2\right) \right] \right\}
\label{qz-fnp}
\end{equation}
where the explicit $\ln(Q^2\,b_{max}^2/c^2)$ dependence is derived
from the evolution equation in Eq.~(\ref{css-W-evo}).  In 
Eq.~(\ref{qz-fnp}), the $b^2$ term represents the leading power
corrections to the renormalization group equations of the kernels $K$ 
and $G$; and the $(b^2)^\alpha$ term is a consequence of extrapolating
the leading power part of the kernels $K$ and $G$ to the small
momentum scale $1/b$ with all powers of running coupling constants
resummed \cite{AMS-DY}.  The actual size of $g_2$ signals the size of
dynamical power corrections.  The choices for the parameters: $g_1$,
$g_2$, and $\alpha < 1$, will be discussed in Sec.~\ref{sec4}.  

We summarize this subsection by writing down our derived extrapolation 
\cite{QZ1} 
\begin{equation}
\tilde{W}^{QZ}(b,Q,x_A,x_B) = \left\{
\begin{array}{ll}
 \tilde{W}(b,Q,x_A,x_B) & \quad \mbox{$b\leq b_{max}$} \\
 \tilde{W}(b_{max},Q,x_A,x_B)\,
 \tilde{F}^{NP}_{QZ}(b,Q,x_A,x_B;b_{max})
                        & \quad \mbox{$b > b_{max}$}
\end{array} \right.
\label{qz-W-sol}
\end{equation}
with the $\tilde{W}(b,Q,x_A,x_B)$ given in Eq.~(\ref{css-W-ij}), and 
the function $\tilde{F}^{NP}_{QZ}$ specified in Eq.~(\ref{qz-fnp}). 
Since the evolution equation in Eq.~(\ref{css-W-evo}) and the
renormalization group equations in Eqs.~(\ref{css-K-rg}) and
(\ref{css-G-rg}) do not include any power corrections, the solution of
these equations, $\tilde{W}(b,Q,x_A,x_B)$ in Eq.~(\ref{qz-W-sol}), is
valid only for $b<b_{c}$ with $b_{c}\sim 0.75$~GeV$^{-1}$, which was
estimated by setting $\ln(1/b_{c}^2) \sim b_{c}^2$. 
Therefore, the numerical value of $b_{max}$ in Eq.~(\ref{qz-W-sol})
should not be larger than the $b_c$ in order to be consistent with
the approximation used to derive the $\tilde{W}(b,Q,x_A,x_B)$.

\subsection{Corrections from parton's intrinsic transverse momentum}
\label{sec3c}

In Eq.~(\ref{qz-fnp}), all $b$-dependence in the $F^{NP}_{QZ}$ 
are dynamical in nature from the way we solve the evolution and
renormalization group equations.  We show in this subsection that
there could be corrections from $Q$-independent intrinsic
$b$-dependence to the $b$-space distribution
$\tilde{W}^{QZ}(b,Q,x_A,x_B)$ in Eq.~(\ref{qz-W-sol}). 

For an arbitrary function $F(b)$, we introduce 
\begin{equation}
W_{ij}(b,Q,x_A,x_B) \equiv F(b)\, \tilde{W}_{ij}(b,Q,x_A,x_B)\, ,
\label{qz-W}
\end{equation}
and find that both $W_{ij}(b,Q,x_A,x_B)$ and
$\tilde{W}_{ij}(b,Q,x_A,x_B)$ can be solutions of the same evolution
equation in Eq.~(\ref{css-W-evo}).  In principle, the function $F(b)$
can also have dependence on parton flavors $i$ and $j$.  If we use the
$W_{ij}(b,Q,x_A,x_B)$ instead of $\tilde{W}_{ij}(b,Q,x_A,x_B)$ as our
solution for the QCD resummed $b$-space distribution in
Eq.~(\ref{css-W-ij}), the function $F(b)$ should be very close to one
when $b$ is small $(<b_{max})$.  Otherwise, its inclusion will not be
consistent with the leading power QCD calculation because the
dominant physics in the small $b$ region has been included in the
perturbatively calculated $\tilde{W}_{ij}(b,Q,x_A,x_B)$.  But,
when $b$ is larger than $b_{max}$, neither the evolution equation nor 
the renormalization group equations have any constraints on the
functional form of the $F(b)$, as long as it is not a function of
$Q$. 

Physically, however, we do not have an arbitrary function $F(b)$.  The
inclusion of any function $F(b)$ should have a correct physics
origin.  In the above derivation of the $\tilde{W}_{ij}(b,Q,x_A,x_B)$,
we added the missing physics (power corrections) to the QCD
resummation formalism when $b>b_{max}$.  However, we did not include
the effect of partons' 
intrinsic transverse momentum, which should appear as a part of the  
$\tilde{W}_{ij}(b,c/b,x_A,x_B)$ --- the boundary condition for the
evolution equation.  When $b$ is small, the factorized formula for the
$\tilde{W}_{ij}(b,c/b,x_A,x_B)$ in Eq.~(\ref{css-W-pert}) should be
reliable.  But, when $b$ is larger than $b_{max}$ or $1/b$ is of the
order of partons' intrinsic transverse momentum, the perturbative QCD
factorized formula requires a so-called ``$Q_T$ smearing'' to be
consistent with experimental data \cite{Owens-RMP}.  A Gaussian-like
smearing function is often used and does a good job in interpreting
the data \cite{Owens-RMP}.  In $b$-space, we can include the effect
due to partons' nonvanishing intrinsic transverse momentum by choosing
\begin{equation}
F(b)= \exp\left(-\bar{g}_2\, b^2\right)\, ,
\label{qz-fb}
\end{equation}
with a constant $\bar{g}_2$, which should be of order $\Lambda_{\rm
QCD}^2$.  Let $\bar{g}_2=\Lambda_{\rm QCD}^2$, we estimate $F(b)\ge
0.99$ for $b<b_{max}$, which gives literally no effect to the
perturbative regime.  However, when $b\gg b_{max}$, the $F(b)$ is
expected to have a sizable effect on the low energy Drell-Yan data.

To include the corrections due to partons' nonvanishing intrinsic
transverse momentum, we modify our extrapolation in
Eq.~(\ref{qz-W-sol}) as follows
\begin{equation}
\tilde{W}^{QZ}(b,Q,x_A,x_B) = \left\{
\begin{array}{ll}
 \tilde{W}(b,Q,x_A,x_B) & \quad \mbox{$b\leq b_{max}$} \\
 \tilde{W}(b_{max},Q,x_A,x_B)\,
 F^{NP}_{QZ}(b,Q,x_A,x_B;b_{max})
                      & \quad \mbox{$b > b_{max}$}
\end{array} \right.
\label{qz-W-sol-m}
\end{equation}
where the perturbatively calculated $\tilde{W}(b,Q,x_A,x_B)$ is
the same as that in Eq.~(\ref{qz-W-sol}).  The modified
nonperturbative function $F_{QZ}^{NP}$ in Eq.~(\ref{qz-W-sol-m}) is
given by 
\begin{eqnarray}
F^{NP}_{QZ}(b,Q,x_A,x_B;b_{max}) 
& = & 
\exp\Bigg\{ -\ln\left(\frac{Q^2 b_{max}^2}{c^2}\right) \left[
    g_1 \left( (b^2)^\alpha - (b_{max}^2)^\alpha\right)
   +g_2 \left( b^2 - b_{max}^2\right) \right] 
\nonumber \\
&\ & {\hskip 0.8in}
   -\bar{g}_2 \left( b^2 - b_{max}^2\right) \Bigg\}\, .
\label{qz-fnp-m}
\end{eqnarray}
Although the terms with $g_2$ and $\bar{g}_2$ have the same
$b$-dependence, they have different physics origins.  The term with
$\bar{g}_2$ represents the effect of partons' nonvanishing intrinsic
transverse momentum.  The term with $g_2$ comes from the
dynamical power corrections, and has an explicit dependence on $Q$.   
In deriving Eq.~(\ref{qz-W-sol-m}), we neglected the intrinsic
transverse momentum corrections $\exp(-\bar{g}_2\, b^2)$ to the
$\tilde{W}$ in the perturbative region, which is consistent with
keeping only the leading power QCD calculations in this region.

Our extrapolation defined in Eq.~(\ref{qz-W-sol-m}) clearly separates
the calculable perturbative region from the large $b$ nonperturbative
region.  In addition, all $b$-dependence in Eq.~(\ref{qz-fnp-m}) have
their own physics origins.  The $(b^2)^\alpha$-dependence mimics the
summation of the perturbatively calculable leading power
contributions to the kernels $K$ and $G$ to all orders in the running
coupling constant $\alpha_s(\mu)$ with the scale $\mu$ running into
the nonperturbative region \cite{AMS-DY}.  The $b^2$-dependence is a
direct consequence of dynamical power corrections to the
renormalization group equations of the kernels $K$ and $G$.
We did not include power corrections to the evolution equation
because of our choice of the $b_{max}$.  We believe that when $Q^2$ is 
much larger than $c^2/b_{max}^2$, our extrapolation defined in
Eq.~(\ref{qz-W-sol-m}) should give a good description of the
$b$-dependence in the region not too much larger than $b_{max}$, which
is most relevant to the heavy boson production.  Uncertainties for not
including higher power corrections 
can be tested by studying the sensitivities on the parameters $g_2$
and $b_{max}$ \cite{QZ1}. 

\subsection{Extrapolation to low $Q^2$}
\label{sec3d}

When we apply our extrapolation defined in Eqs.~(\ref{qz-W-sol-m}) and
(\ref{qz-fnp-m}) to the low $Q^2$ Drell-Yan data, we might need a few
modifications.  As discussed in Sec.~\ref{sec2}, the Fourier transform
from $b$-space to $Q_T$-space is sensitive to the large $b$ region
when $Q^2$ and $\sqrt{S}$ are both small.  Therefore, the $Q_T$
distributions at low energies are much more sensitive to the
parameters $g_2$ and $b_{max}$ or even higher power corrections.  

For deriving our extrapolation, we systematically dropped the power
corrections of momentum scale between $1/b_{max}^2$ to $Q^2$.  
When $Q^2$ is small, the leading power perturbative QCD calculations
receive relatively larger $1/Q^2$ corrections, in particular, the
type-two power corrections.  For example, a $1/Q^2$ term in the
evolution equation in Eq.~(\ref{css-W-evo}) results into a type-two 
power correction to the perturbatively calculated
$\tilde{W}(b,Q,x_A,x_B)$ in Eq.~(\ref{qz-W-sol-m}).  The size of the
type-two power corrections resummed from $\Lambda^2/Q^2$ to
$\Lambda^2/(1/b_{max}^2)$ is proportional to the numerical value of
$b_{max}^2\Lambda^2$, where $\Lambda^2$ is the characteristic scale of
the power corrections.  For a larger $b_{max}$, the perturbatively
calculated $\tilde{W}(b,Q,x_A,x_B)$ receives a larger power
correction.  

On the other hand, the leading power contributions are resummed from 
$\ln(1/b_{max}^2)$ to $\ln(Q^2)$, and are proportional to the
numerical value of $\ln(Q^2)$.  When $Q^2\sim M_Z^2$ at collider
energies, the $\ln(Q^2)$ is much larger than the $b_{max}^2
\Lambda^2$, and the power corrections to the perturbatively
calculated $\tilde{W}(b,Q,x_A,x_B)$ can be neglected.  However, at the
fixed target energies, both $Q^2$ and the leading power contributions
are much smaller.  Therefore, the power corrections to the
perturbatively calculated leading power contributions become
relatively more important at fixed target energies.  

Therefore, in order to reduce the relative size of the power
corrections in the perturbative (or small $b$) region, we need to
reduce the numerical value of 
$b_{max}$. On the other hand, we prefer to keep the $b_{max}$ as
large as possible to have more contributions from the perturbative
region in order to have more prediction than parameter
fitting.  The $b_{max}=0.5$~GeV$^{-1}$ was proposed in
Ref.~\cite{CSS-W}.  For our numerical results in Sec.~\ref{sec4}, we
will test the sensitivities on the choices of $b_{max}$.  

The $\ln(Q^2\,b_{max}^2/c^2)$ dependence in our $F^{NP}_{QZ}$ is a
direct consequence of dropping the type-two power corrections to the
evolution equation in Eq.~(\ref{css-W-evo}). Therefore, in
order to preserve the $\ln(Q^2\,b_{max}^2/c^2)$ dependence in our
extrapolation, we expect to require a smaller $b_{max}$ for a better
description of the low energy Drell-Yan data. 


\section{Numerical Results and Conclusions} 
\label{sec4}

In this section, we numerically compare the $Q_T$ distributions
derived from the $b$-space resummation formalism with experimental
data from $W$ and $Z$ production at collider energies to the low
energy Drell-Yan processes.

\subsection{Numerical accuracy}
\label{sec4a}

One of the potential drawbacks of the $b$-space resummation formalism
is the difficulty of matching the resummed and fixed-order predictions
to the $Q_T$ distributions at large $Q_T$ \cite{QCD-rpt}.  It was
generally believed \cite{QCD-rpt} that the $b$-space resummed $Q_T$ 
distribution from Eq.~(\ref{css-gen}) becomes unphysical or negative
when $Q_T$ is large.  For example, a matching between the resummed and
fixed-order calculations has to take place at $Q_T \sim 50$~GeV for
$W$ production when these two predictions cross over \cite{Ellis-2}.
On the other hand, as we discussed in Sec.~\ref{sec2}, we expect 
the predictions derived from the $b$-space resummation formalism in
Eq.~(\ref{css-gen}) to work better when $Q_T$ is larger because (1)
the $b$-integral is dominated by the smaller $b$ region and (2) the 
perturbatively calculated $Y$ term is larger than the resummed
$\tilde{W}$ term.  We find that this puzzle was mainly caused by
the lack of the numerical accuracy of the Bessel function used to
perform the Fourier transform in Eq.~(\ref{css-W-F}).  As we show
below, the $Q_T$ distributions derived from the $b$-space resummation
formalism are smoothly consistent with data for all transverse momenta
up to $Q$. 

Due to the oscillatory nature of the Bessel function, high
numerical accuracy of the $J_0(z)$ with $z=Q_T\,b$ in
Eq.~(\ref{css-W-F}) is necessary for ensuring an accurate cancellation
in large $z$ region for a reliable $b$-integration.  Because
$z$ is proportional to $Q_T$, the number of oscillations of the Bessel
function strongly depends on the value of $Q_T$ for the same range of
$b$.  For example, when $b \in (0, 2)$~GeV$^{-1}$, $J_0(Q_T b)$
crosses zero 0, 6, and 63 times for $Q_T=1$, 10, and 100~GeV,
respectively.  It is clear that numerical accuracy of the Bessel
function is extremely important for the large $Q_T$ region.   We
noticed that most work published in the literature used some kind of
asymptotic form to approximate the Bessel function when $z=Q_T b$ is
large. We find that the usage of an asymptotic form for the Bessel
function is a major source of the uncertainties observed for the large
$Q_T$ region.  Instead of using an asymptotic form, we use the
following integral form for the Bessel function 
\begin{equation}
J_0(z) = \frac{1}{\pi}\, 
\int_0^\pi \cos\left(z\sin(\theta)\right) d\theta\, .
\label{bessel}
\end{equation}
The great advantage of using an integral form is that we can control
the numerical accuracy of the Bessel function by improving the
accuracy of the integration in Eq.~(\ref{bessel}).  One can test the
numerical accuracy of the Fourier transform in Eq.~(\ref{css-W-F}) by
using functions whose Fourier transform can be carried out
analytically. In view of the nonperturbative $b$-dependence in large
$b$ region, we used two functions: $\exp(-\sigma\, b)$ and
$\exp(-b^2/\sigma^2)$ to test the numerical accuracy of the
$b$-integration in Eq.~(\ref{css-W-F}).  The $b$-integration for these
two functions can be carried out analytically.  Having the analytical
solution in $Q_T$-space, we can study the convergence and the 
numerical accuracy of the Fourier transform at different $Q_T$ by
varying the parameter $\sigma$ of these two functions.  We find that
by using the integral form of the Bessel function, the numerical
integration over $b$ defined in Eq.~(\ref{css-W-F}) is very accurate
for a wide range of $\sigma$ and $Q_T$, and the accuracy is only
limited by the precision of variables used in a computer programming
language.  

\subsection{$Q_T$ distributions of $W$ and $Z$ production}
\label{sec4b}

In this subsection, we compare the predictions of the $b$-space
resummation formalism with Fermilab data on $W$ and $Z$ production, 
and quantitatively demonstrate the excellent predictive power of the
$b$-space resummation formalism at collider energies \cite{QZ1}.  We
show that the $b$-space resummed $Q_T$ distributions are very
insensitive to the parameters in the nonperturbative $F^{NP}_{QZ}$.

Since we are not interested in the detailed fitting to the data in
this paper, we did not perform any simulation on final-state cuts to
improve the theory curves in the following plots.  In all plots,
CTEQ4M parton distributions are used; and for the $Y$ term in
Eq.~(\ref{css-gen}), we use $\frac{1}{2}\sqrt{Q^2+Q_T^2}$ for the
factorization and renormalization scale $\mu$ in Eq.~(\ref{css-W-yn}).
For $W$ and $Z$ production,
a fixed range of the rapidity was integrated and a narrow width
approximation was used for $Q^2$ integration \cite{pink-book}.  

\begin{figure}
\epsfig{figure=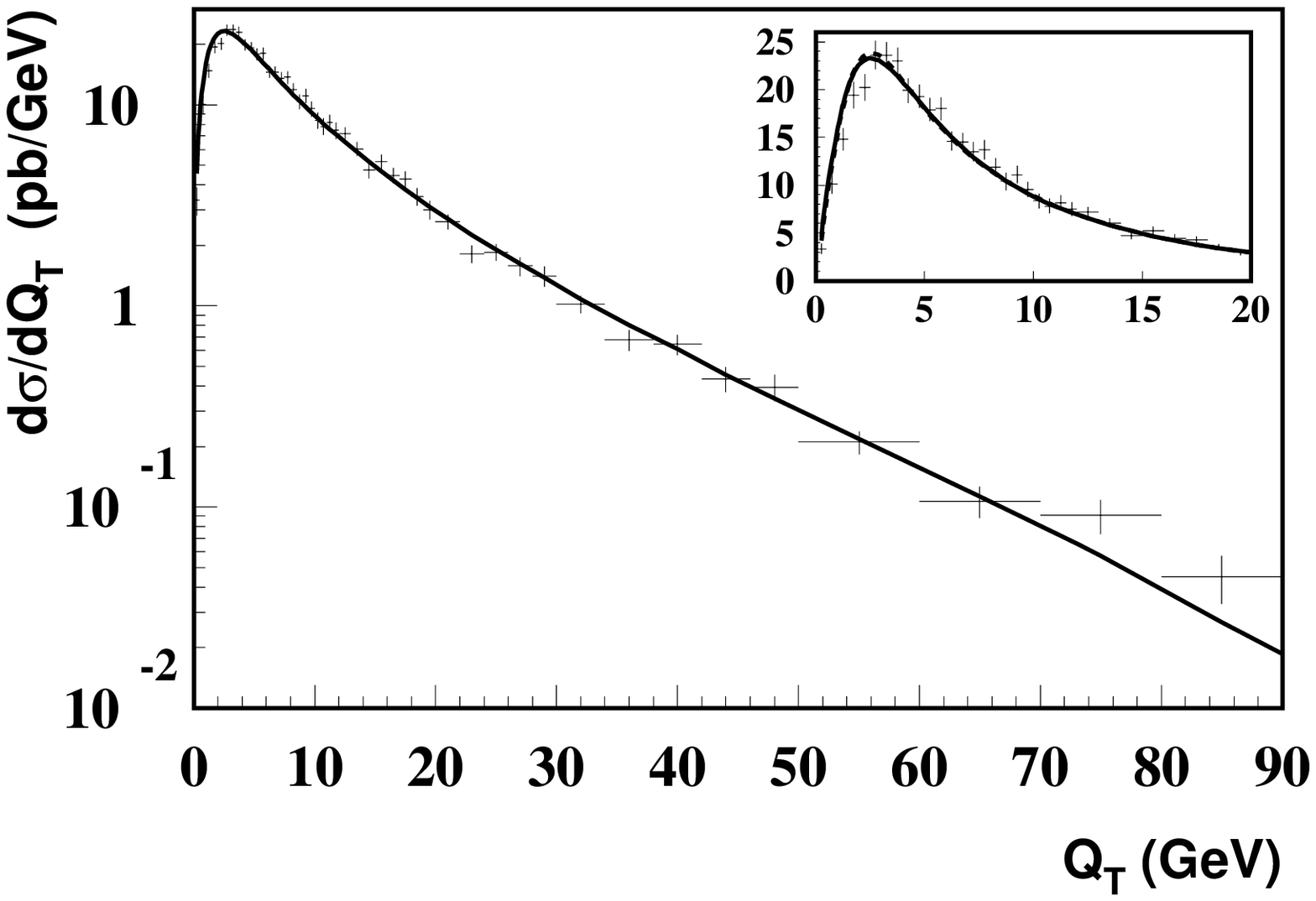,width=6.0in}
\caption{Comparison between the $b$-space resummed $Q_T$ distribution
and CDF data \protect\cite{CDF-Z}. The inset shows the $Q_T<20$~GeV
region.} 
\label{fig7}
\end{figure}

\begin{figure}
\epsfig{figure=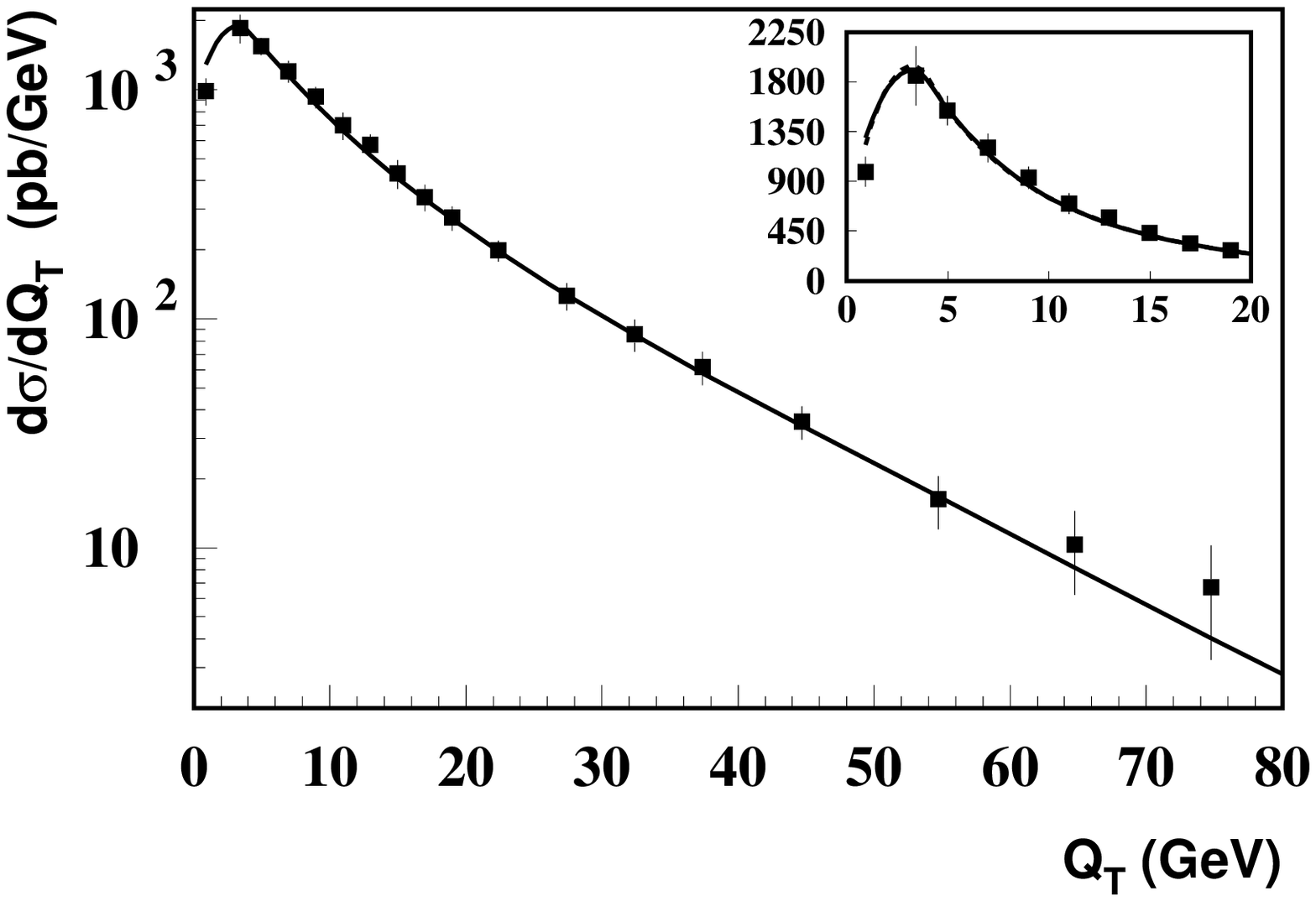,width=6.0in}
\caption{Comparison between the $b$-space resummed $Q_T$ distribution
and D0 data \protect\cite{D0-W}.  The inset shows the $Q_T<20$~GeV
region.}
\label{fig8}
\end{figure}

We test the sensitivities on the parameters in the $F^{NP}_{QZ}$ by 
first setting $g_2=0$ and $\bar{g}_2=0$ (no ``power'' corrections).
We then fix $g_1$ and $\alpha$ in Eq.~(\ref{qz-fnp-m}) by requiring
the first and second order derivatives of the $\tilde{W}$ to be
continuous at 
$b=b_{max}=0.5$~GeV$^{-1}$, and plot our predictions (solid lines) to
the $Q_T$ distributions of $Z$ and $W$ production at Tevatron in
Figs.~\ref{fig7} and \ref{fig8}, respectively.  In Fig~\ref{fig7}, we
plot the $d\sigma/dQ_T$ of $e^+e^-$ pairs as a function of $Q_T$ at
$\sqrt{S}=1.8$~TeV.  The data are from CDF Collaboration \cite{CDF-Z}.
Theory curves (Z-only) are from Eq.~(\ref{css-gen}) with the
$\tilde{W}$ given in Eq.~(\ref{qz-W-sol-m}).  Same as in
Ref.~\cite{CDF-Z}, an overall normalization 1.09 was used. 
In Fig.~\ref{fig8}, we plot the $d\sigma/dQ_T$ for $W$ production with
the same $b_{max}$ and $g_2$.  The data for $W$ production are from D0
Collaboration \cite{D0-W}.  For the theory curves, we integrate the
rapidity of $W$ particle from -2 to +2 and set the overall
normalization to be one.  From Figs.~\ref{fig7} and \ref{fig8}, it is
clear that the QCD predictions from the $b$-space resummation
formalism are consistent with the data for all $Q_T<Q$.  

We notice that the theory curves in Figs.~\ref{fig7} and \ref{fig8}
are slightly below the data at large $Q_T$.  We believe that it is
because we have only the leading order contribution 
to the $Y$ term in Eq.~(\ref{css-gen}).  At large $Q_T$, the $Y$ term
dominates.  Similar to the fixed-order perturbative calculations, the
next-to-leading order contribution will enhance the theoretical
predictions \cite{AER}.  In addition, the inclusion of virtual photon
channel and its interference with $Z$-channel should shift the peak of
the theory curves to a slightly larger $Q_T$ value and make the theory
curves closer to the data \cite{Ellis-1,CDF-Z}. 

We now test the theory curves' sensitivities on the parameters of
$F^{NP}_{QZ}$.  Since $Q$ is fixed to the mass
of the vector boson and the terms with $g_2$ and $\bar{g}_2$ have the
same $b$-dependence, we can simplify the following discussion by
rewriting the $F^{NP}_{QZ}$ in Eq.~(\ref{qz-fnp-m}) for $W$ and $Z$
production as follows \cite{QZ1}, 
\begin{equation}
F^{NP}_{WZ}(b,Q,x_A,x_B;b_{max}) =
\exp\bigg\{ - g_1 \left( (b^2)^\alpha - (b_{max}^2)^\alpha\right)
            - g_2 \left( b^2 - b_{max}^2\right) \bigg\}.
\label{qz-fnp-wz}
\end{equation}
We let $g_2$ in Eq.~(\ref{qz-fnp-wz}) be a fitting parameter for
any given value of $b_{max}$ and fix $g_1$ and $\alpha$ by the
derivatives.  Although the fitting prefers $g_2\sim 0.8$~GeV$^{2}$,
the $Q_T$ distributions are extremely insensitive to the choices of
$b_{max}$ and $g_2$.  The total $\chi^2$ are very stable for $b_{max}
\in (0.25, 0.8)$~GeV$^{-1}$ and $g_2 \in (0, 1.6)$~GeV$^{2}$.  In
Figs.~\ref{fig7} and \ref{fig8}, we also plot the theory curves
(dashed lines) with $g_2=0.8$~GeV$^{2}$ (the best fitting value).
Non-vanishing $g_2$ gives a small improvement to the $Q_T$
distributions at small $Q_T$.  We then vary the value of $\alpha$ in
Eq.~(\ref{qz-fnp-wz}) by requiring only the first order derivative to
be continuous at $b=b_{max}$, and find equally good theoretical
predictions, except very mild oscillations in the curves at very large
$Q_T$ due to the Fourier transform of a less smoother $b$-space
distribution.  It is well-known that when $Q_T$ is larger, 
any small kink in the $b$-space distributions transforms into
oscillations in the $Q_T$ distributions due to more rapid oscillations
from the Bessel function $J_0(Q_T\,b)$.  The observed insensitivity
on $b_{max}$, $g_2$, and $\alpha$ is a clear evidence that the
$b$-space resummation formalism is not sensitive to the power
corrections at collider energies.  That is, at collider energies, a
direct extrapolation of the leading power contributions to the large
$b$ region (the $(b^2)^\alpha$ term) represents the most relevant
$b$-dependence at large $b$. 

\begin{figure}
\begin{center}
\epsfig{figure=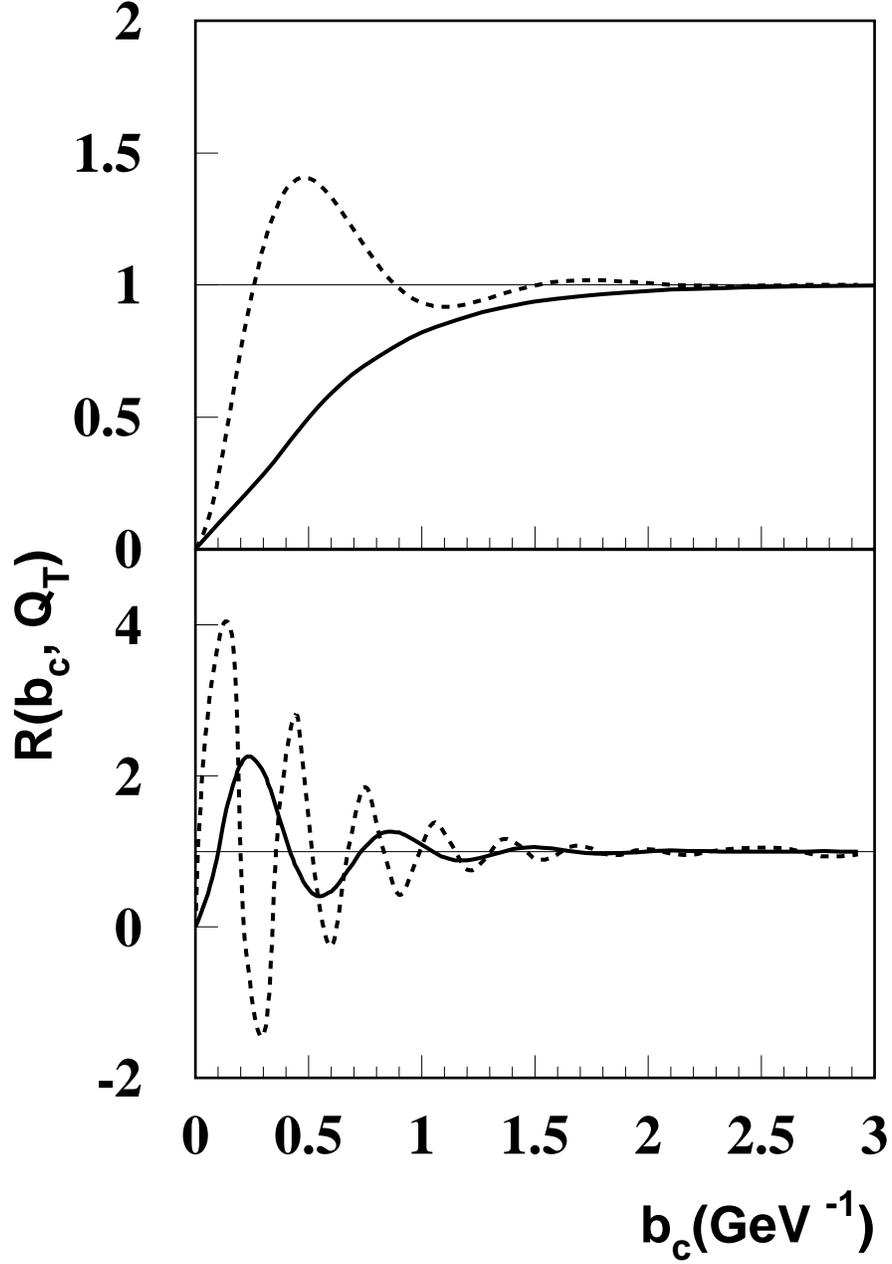,width=4.6in}
\end{center}
\caption{Ratio $R(b_c,Q_T)$ in Eq.~(\protect\ref{W-bc}) as a function
of $b_c$ for $Z$ production at Tevatron energy: $Q_T=0$~GeV
(solid), $Q_T=5$~GeV (dashed) in the top plot; $Q_T=10$~GeV (solid),
and $Q_T=20$~GeV (dashed) in the bottom plot.}   
\label{fig9}
\end{figure}

To further test the sensitivities on the nonperturbative $F^{NP}$, we
introduce a cutoff $b_{c}$ to the $b$-integration in 
Eq.~(\ref{css-W-F}),
\begin{equation}
w(b_c, Q_T) \equiv
\frac{1}{2\pi}\int_0^{b_c} db\, b\, J_0(Q_T b)\, 
\tilde{W}^{QZ}(b,Q,x_A,x_B)\, ,
\label{css-bc}
\end{equation}
and define the ratio,
\begin{equation}
R(b_c,Q_T) \equiv \frac{w(b_c,Q_T)}{w(b_c=\infty,Q_T)}\, .
\label{W-bc}
\end{equation}
Because of the oscillatory nature of the Bessel function, we expect the
ratio $R(b_c,Q_T)$ to oscillate around one and eventually
converge to one as $b_c$ increases.  In Fig.~\ref{fig9}, we plot the
ratio $R(b_c,Q_T)$ as a function of $b_c$ for $Z$ production at
different values of $Q_T$.  In plotting the theory curves in
Fig.~\ref{fig9}, we set $\sqrt{S}=1.8$~TeV, $b_{max}=0.5$~GeV$^{-1}$,
and $g_2=0.8$~GeV$^2$.  For $Q_T=0$, 5, 10, and 20~GeV, as shown in
Fig.~\ref{fig9}, the $R(b_c,Q_T)$ quickly reaches one 
at $b_c\sim 1/Q_T$ as $b_c$ increases.  Even at $Q_T=0$,
$R(b_c,Q_T)$ is close to one within a few percents at $b_c\sim
2$~GeV$^{-1}$.  It tells us that the physics of the $Q_T$ distribution
is dominated by the perturbative (or small $b$) region.  However, due
to the oscillatory nature of the Bessel function, the $b$-integral for
the Fourier transform in Eq.~(\ref{css-gen}) converges roughly
at a common value of $b_c\sim 2$~GeV$^{-1}$, which is larger than
$b_{max}$.  That is, the nonperturbative extrapolation to the large
$b$ region is necessary to ensure the correct cancellations in the
large $b$ region.  The small dependence on the parameters $g_2$ and
$b_{max}$ shows that our extrapolation defined in Eq.~(\ref{qz-fnp-m})
catches the most of the physics in this region, and higher power
corrections are not important.

In addition, we show quantitative $g_2$-dependence of the $b$-space
resummed $Q_T$ distributions by introducing the following ratio,
\begin{equation}
R_{\sigma}(Q_T,g_2) \equiv \left.
\frac{d\sigma(g_2)}{dQ^2 dy dQ_T^2} \right/
\frac{d\sigma(g_{2_b})}{dQ^2 dy dQ_T^2}\, , 
\label{Sigma-g2}
\end{equation}
where $d\sigma/dQ^2 dy dQ_T^2$ is defined in Eq.~(\ref{css-gen}), and
$g_{2_b}=0.8$~GeV$^2$ is the best fit value for $g_2$.  In
Fig.~\ref{fig10}, we plot the ratio $R_{\sigma}(Q_T,g_2)$ in
Eq.~(\ref{Sigma-g2}) as a function of $Q_T$ for $Z$ production at the
Tevatron energy.  The solid line corresponds to $g_2=2\, g_{2_b}$,
while the dashed line has the $g_2=\frac{1}{2}\, g_{2_b}$.  For almost
all $Q_T<Q$, the deviation of the $b$-space resummed $Q_T$
distributions at the Tevatron energy is less than two percents when we
change $g_2$ from half to twice of the best fit $g_2$ value.  The
small irregularity for $Q_T>50$~GeV is due to 
the fact that the calculated cross section has about half of a percent
numerical uncertainty in this region, which is of the same order as
the deviation.  We find that the deviation at the LHC energy is much
less than one percent.  That is, the Fourier transform from $b$-space
to $Q_T$-space is completely determined by the perturbatively
calculated $b$-space distribution plus our extrapolation, which is
totally fixed by the continuity at $b=b_{max}$.  Therefore, we
conclude that the $b$-space resummed $Q_T$ distributions for vector
boson productions at collider energies have an excellent predictive
power. 

\subsection{Drell-Yan $Q_T$ distributions at low energies}
\label{sec4c}

As discussed in Sec.~\ref{sec2}, the Drell-Yan $Q_T$ distributions at
fixed target energies are much more sensitive to the nonperturbative
input at large $b$.  The predictive power of the $b$-space resummation
formalism may be limited.  To explore the predictive power of the
$b$-space resummed $Q_T$ distributions for the low energy Drell-Yan
process, we compare the low energy Drell-Yan data with the predictions
from the CSS $b$-space resummation formalism plus the extrapolation
defined in Eq.~(\ref{qz-W-sol-m}).  We 
use a subset of available data to fix the parameters in the
nonperturbative $F^{NP}_{QZ}$, and then, we compare
the predictions calculated by using these parameters with the rest
of the data. 

When $Q_T\leq 4$~GeV, the Bessel function does not have any
oscillation in the perturbative region ($b<b_{max}$).  A less
smoother connection of the $\tilde{W}(b,Q,x_A,x_B)$ at $b=b_{max}$
does not produce any apparent oscillation for the $Q_T$ distribution
in this region.  Therefore, we do not have to require the derivative
of the $\tilde{W}^{QZ}(b,Q,x_A,x_B)$ in Eq.~(\ref{qz-W-sol-m}) to be
continuous at $b=b_{max}$.  In this subsection, we treat $g_1$, $g_2$,
$\bar{g}_2$, and $\alpha$ in the $F^{NP}_{QZ}$ as
fitting parameters.  In order to maximize the predictive power of the
$b$-space resummed $Q_T$ distributions, we try to fit the data
with the least number of parameters, and then add more parameters to
see the improvements.  By going through this fitting procedure, we can 
learn the importance of each parameter and the corresponding physics.

We choose the following Fermilab data to fit the parameters in the
$F^{NP}_{QZ}$: $Q\in (5,6)$~GeV and $Q\in (7,8)$~GeV from E288
\cite{DY-E288}; and $Q\in (7,8)$~GeV and $Q\in (10.5,11.5)$~GeV from
E605 \cite{DY-E605}.  We have a cut on the transverse momentum:
$Q_T\le 1.4$~GeV, which is the same as that used in Ref.~\cite{LBLY}.
Since the $g_2$ dependence is very small at collider 
energies, we try to fit these Drell-Yan data with two parameters:
$g_1$ and $\alpha$ plus two overall normalizations for two
experiments. Using the $F^{NP}_{QZ}$ in Eq.~(\ref{qz-fnp-m}) and
$b_{max}=0.5$~GeV$^{-1}$, we obtain a convergent fit with 
a total $\chi^2=78$ for 28 data points.  The corresponding numerical
values of the fitting parameters: $g_1$, $\alpha$, and two
overall normalization constants for these two experiments are given in
Table~\ref{table}.  The large $\chi^2$ clearly indicates that power
corrections (or terms proportional to $b^2$ in the $F^{NP}_{QZ}$)
are very important for understanding the low energy data.

We now include the $b^2$ dependence of the $F^{NP}_{QZ}$ into the
fit.  In order to use a minimum number of fitting parameters, we
first combine both $g_2$ and $\bar{g}_2$ terms and define 
\begin{equation}
g_2\,\ln\left(\frac{Q^2\,b_{max}^2}{c^2}\right)
+\bar{g}_2 \equiv g \, .
\label{g2-g}
\end{equation}
We approximate the $g$ as a constant fitting parameter.  With three
fitting parameters: $g_1$, $\alpha$, and $g$ plus two overall
normalizations for two experiments, we fit the same 28 data points
for $b_{max}=0.5$~GeV$^{-1}$.  We obtain a much better fit with a
total $\chi^2=32$, and the corresponding fitting parameters are given
in Table~\ref{table}.  Using $b_{max}=0.5$~GeV$^{-1}$, and the
numerical values of the fitting parameters in the Table~\ref{table},
we find that the $b^2$ term in the $F^{NP}_{QZ}$ is less than the
$(b^2)^\alpha$ term for $b$ as large as $b\sim 2.4$~GeV$^{-1}$ ($\sim
2.0$~GeV$^{-1}$ for $b_{max}=0.3$~GeV$^{-1}$).  It indicates that
although the power corrections (the $b^2$ dependence) are important, 
the extrapolation from the leading power contributions (the
$(b^2)^\alpha$ dependence in the $F^{NP}_{QZ}$) is crucial for
understanding the low energy data.

As we discussed in Sec.~\ref{sec3d}, we can test the size of power
corrections to the perturbative region ($b<b_{max}$) by studying the
$b_{max}$ dependence.  We find that the fitting parameter $\alpha$ is
extremely stable when we change the $b_{max}$, and it prefers a
numerical value around $\alpha\sim 0.15$.  This can be understood as
follows.  The parameter $\alpha$ was introduced in Sec.~\ref{sec3b} to
approximate the $\mu^2$-dependence when we extrapolate the
leading power contributions to the $S(b,Q)$ into the small $\mu^2$
region \cite{AMS-DY}.  Therefore, the parameter $\alpha$ should only
depend on the effective anomalous 
dimensions (resummed to all orders in the running coupling constant,
$\alpha_s(\mu)$), and should not be sensitive to the numerical value
of the $b_{max}$.  We then fix
$\alpha=0.15$ and re-fit those 28 data points with two parameters:
$g_1$ and $g$ plus two overall normalization constants.  

We find a smooth reduction of the total $\chi^2$ when we decrease the
$b_{max}$ to as low as 0.3~GeV$^{-1}$.  For $b_{max}=0.5$, 0.4, and
0.3, we obtain convergent fits with the total $\chi^2=32$, 27, and 24,
respectively.  The corresponding fitting parameters are listed in
Table~\ref{table}.  Although the total $\chi^2$ is still very stable
when we further reduce the $b_{max}$, we stop at
$b_{max}=0.3$~GeV$^{-1}$ because it is difficult to
distinguish the perturbative prediction from the parameter fitting
when $b_{max}$ is too small.  We confirm from the fitting results
listed in Table~\ref{table} that the $(b^2)^\alpha$ term in our
extrapolation is very important and it dominates the transition region
between the perturbative calculation and the nonperturbative
extrapolation.  We learn that the overall normalizations, which are
needed to fit the different experimental data sets, have a strong
dependence on the $b_{max}$.  As shown in Table~\ref{table}, the
overall normalizations are driven to the unity as the $b_{max}$
decreases.  That is, if we reduce the relative size of the power
corrections in the perturbative region ($b<b_{max}$) by 
a reduction of $b_{max}$, both data sets
used in the fit are in an excellent agreement with each other.

\begin{figure}
\epsfig{figure=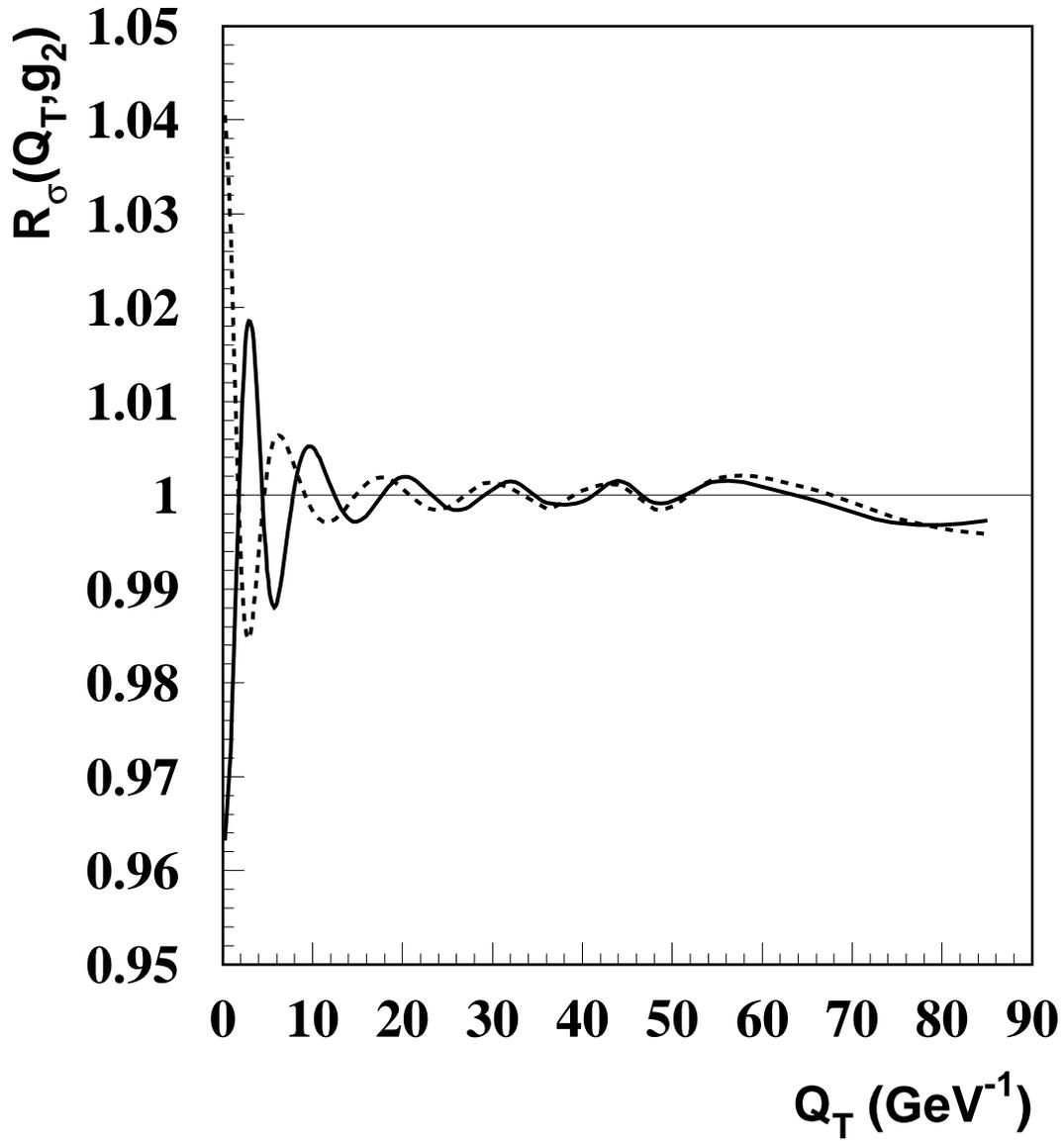,width=5.5in}
\caption{Ratio $R_{\sigma}(Q_T,g_2)$ in
Eq.~(\protect\ref{Sigma-g2}) at $Q=M_Z$ and $\protect\sqrt{S}=1.8$~TeV
as a function of $Q_T$: $g_2=2\, g_{2_b}$ (solid) 
and $g_2=(1/2) g_{2_b}$ (dashed).} 
\label{fig10}
\end{figure}

\begin{figure}
\epsfig{figure=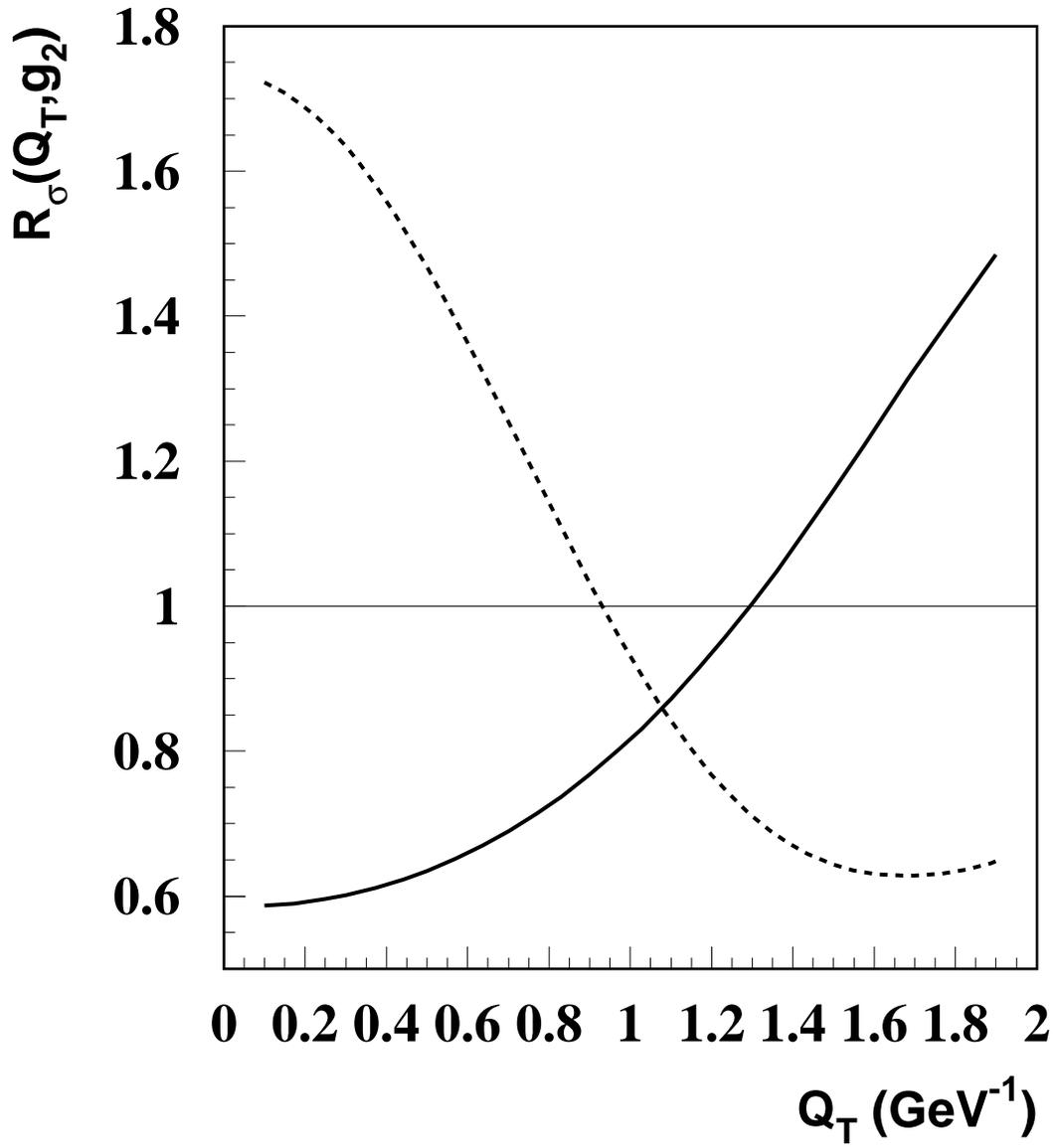,width=5.5in}
\caption{Same as that in Fig.~\protect\ref{fig10} at $Q=6$~GeV and
E288 collision energy.}
\label{fig11}
\end{figure}

In order to show the quantitative size of the power corrections
(the $b^2$ term in the $F^{NP}_{QZ}$) at the low
energy, we replot Fig.~\ref{fig10} at $Q=6$~GeV and
$\sqrt{S}=27.4$~GeV in Fig.~\ref{fig11}.  In plotting
Fig.~\ref{fig11}, we use the ratio $R_{\sigma}(Q_T,g_2)$ defined in
Eq.~(\ref{Sigma-g2}) with the $g_2$ dependence replaced by the $g$
dependence defined in Eq.~(\ref{g2-g}).  The $g_{2_b}$ in
Eq.~(\ref{Sigma-g2}) is replaced by the best fit value $g_b$.   
We choose $b_{max}=0.3$~GeV$^{-1}$ and the corresponding
fitting parameters: $\alpha$, $g_1$, and $g$ given in
Table~\ref{table}.  The best fit value $g_b=0.28$~GeV$^2$.   
In Fig.~\ref{fig11}, the solid line corresponds to $g=2\, g_b$ while
the dashed line has $g=\frac{1}{2}\, g_b$.  For $Q_T$ from 0 to
2~GeV, the $b^2$-dependence can change the $b$-space resummed $Q_T$
distributions by as much as 80 percent when the parameter $g$ changes
from half to twice of the best fit value.  In contrast to the 
two percents deviation at collider energies, as shown in
Fig.~\ref{fig10}, the large variation shown in Fig.~\ref{fig11}
further confirms that the power corrections are very important 
for describing the Drell-Yan data at the fixed target energies
\cite{DY-GQZ}.  

\begin{figure}
\epsfig{figure=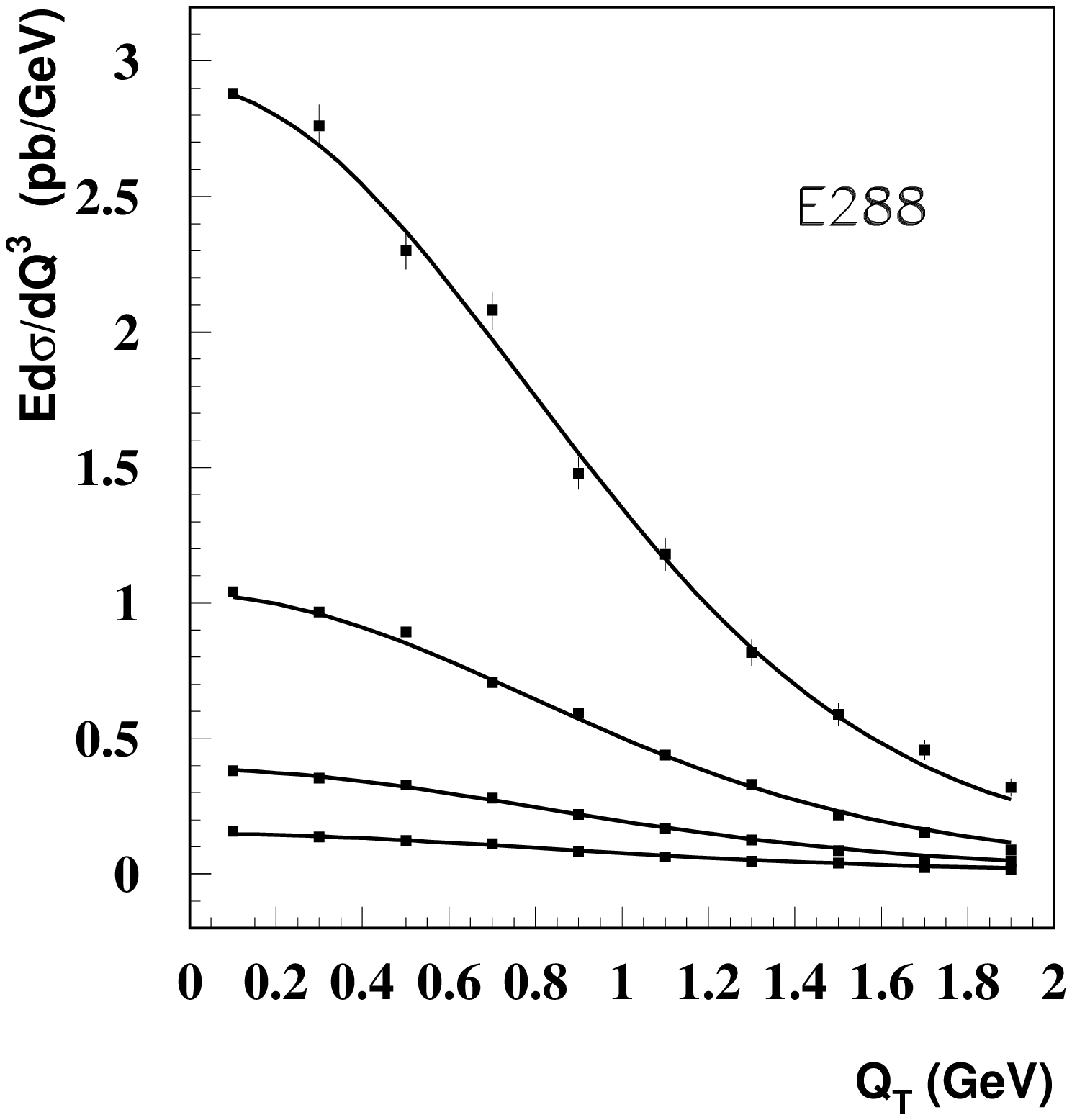,width=5.5in}
\caption{Comparison between the $b$-space resummed $Q_T$ distribution
and Fermilab E288 data \protect\cite{DY-E288}.  The overall
normalization for the theory curves: N$_{\rm E288}=0.97$.}
\label{fig12}
\end{figure}

\begin{figure}
\epsfig{figure=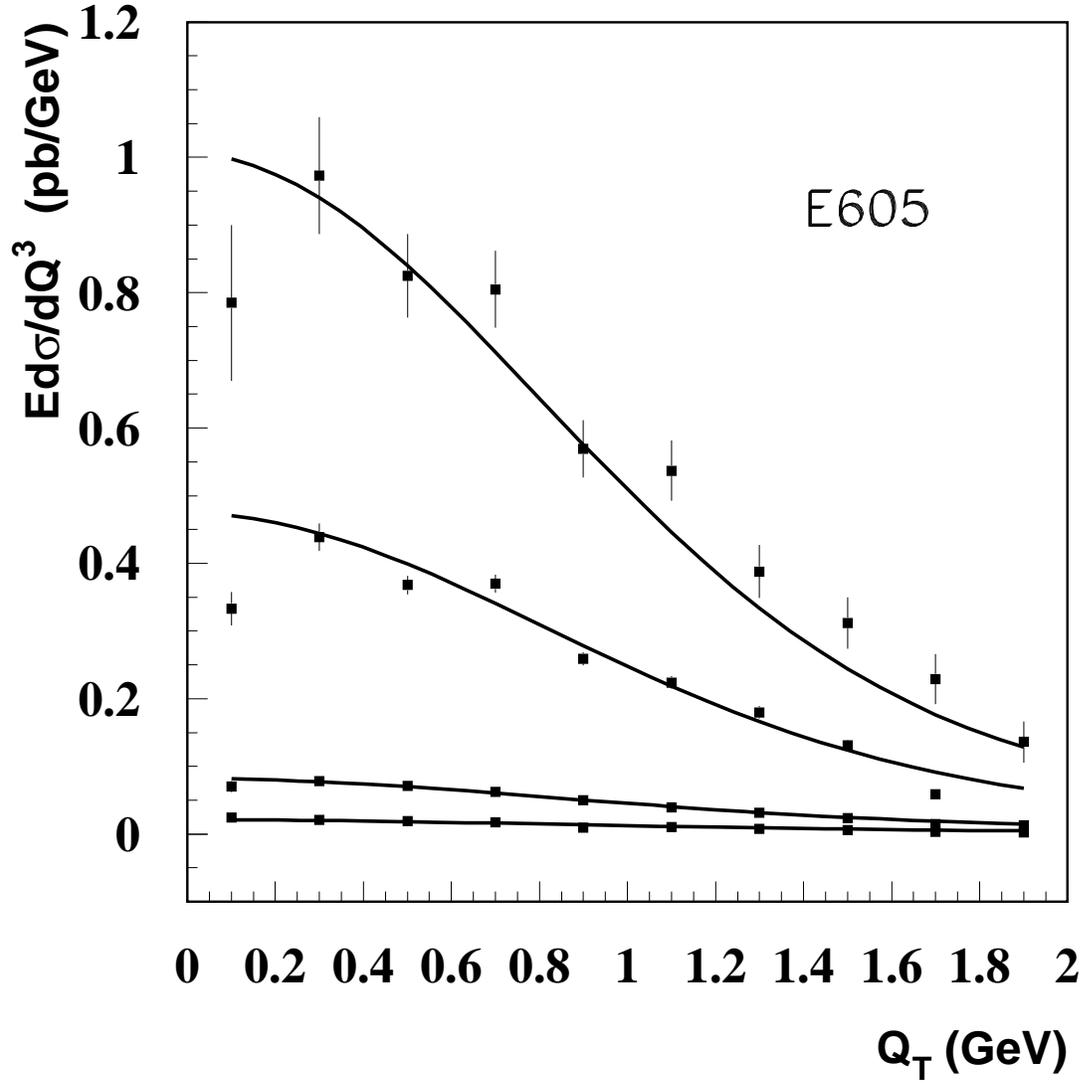,width=5.5in}
\caption{Comparison between the $b$-space resummed $Q_T$ distribution
and Fermilab E605 data \protect\cite{DY-E605}. The overall
normalization for the theory curves: N$_{\rm E605}=1.0$.}
\label{fig13}
\end{figure}

\begin{figure}
\epsfig{figure=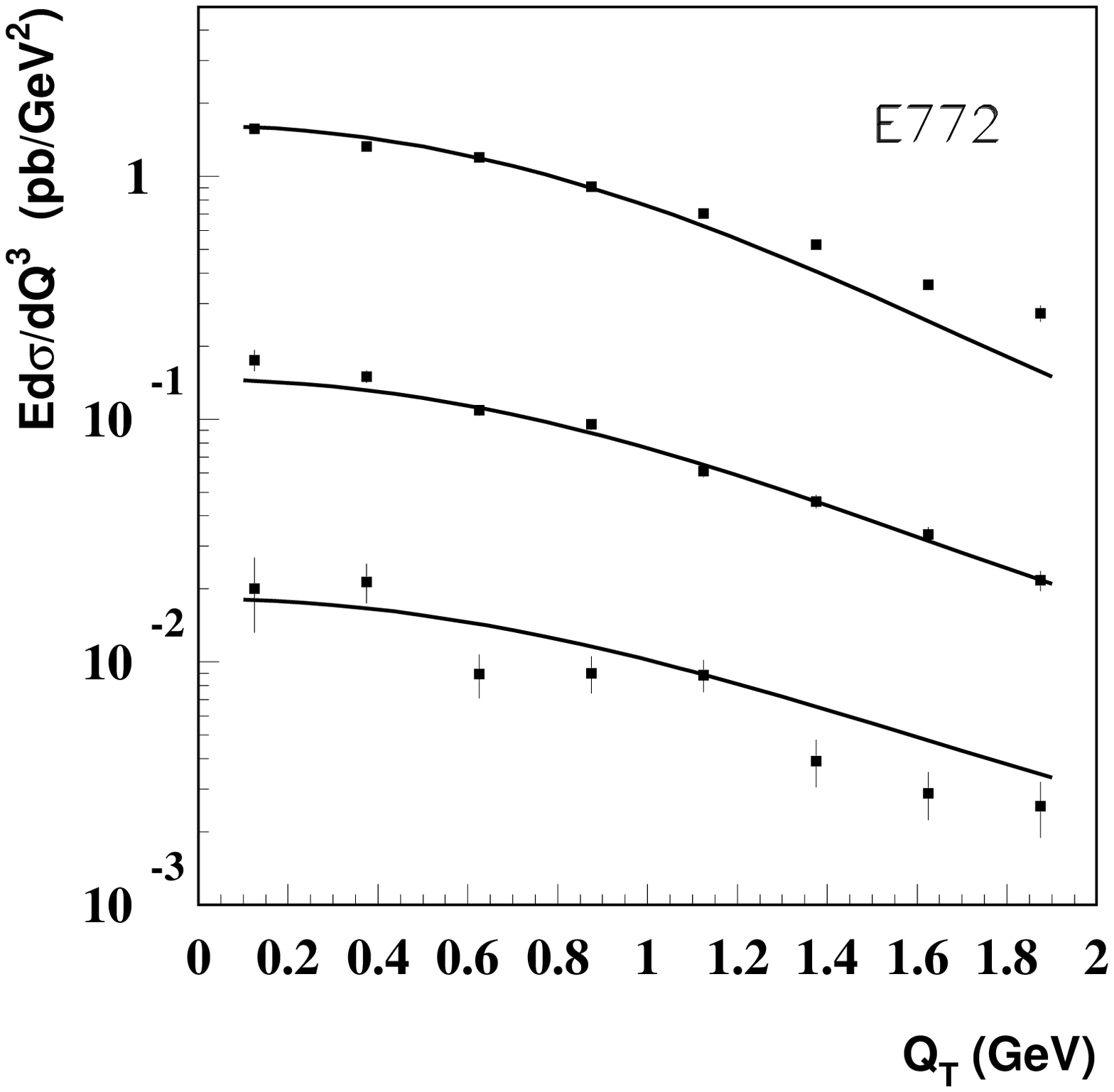,width=5.5in}
\caption{Comparison between the $b$-space resummed $Q_T$ distribution
and Fermilab E772 data \protect\cite{DY-E772}. The overall
normalization for the theory curves: N$_{\rm E772}=1.6$.}
\label{fig14}
\end{figure}

To explore the predictive power of the $b$-space resummed $Q_T$
distributions at fixed target energies, we compare the resummed $Q_T$
distributions with existing Drell-Yan data in Figs.~\ref{fig12},
\ref{fig13}, and \ref{fig14}.  All theory curves in these figures are
calculated with our extrapolation defined in Eq.~(\ref{qz-W-sol-m})
and $b_{max}=0.3$~GeV$^{-1}$.  The nonperturbative extrapolation to
the large $b$ region: $F^{NP}_{QZ}$ is defined in Eq.~(\ref{qz-fnp-m})
with the $g_2$ and $\bar{g}_2$ terms combined as in Eq.~(\ref{g2-g}).
We fix the parameter $\alpha$ to be 0.15, and use the fitted values
$g_1=0.92$~GeV$^{\alpha}$ and $g=0.28$~GeV$^2$ from
Table~\ref{table}. Although we only used 28 data points with $Q_T\le
1.4$~GeV in our fits for determining the two parameters: $g_1$ and $g$,
we plot both theory curves and data for a much enlarged phase space in
Figs.~\ref{fig12}, \ref{fig13}, and \ref{fig14}.  We plot data with
$Q_T$ as large as 2~GeV at different values of $Q$ in order to explore
the predictive power of the theoretical calculations.  In
Fig.~\ref{fig12}, we compare the theoretical calculations with
Fermilab E288 data at $\sqrt{S}=27.4$~GeV \cite{DY-E288}.  The theory
curves are multiplied by an overall normalization constant 
N$_{\rm E288}=0.97$ as listed in Table~\ref{table}, which is 
different from what was found in Ref.~\cite{LBLY}.  From top to bottom
in Fig.~\ref{fig12}, the four curves along with four data sets
correspond to $Q\in (5,6)$~GeV, $(6,7)$~GeV, $(7,8)$~GeV, and
$(8,9)$~GeV, respectively. Two of the four data sets: $Q\in (5,6)$~GeV
and $(7,8)$~GeV with $Q_T<1.4$~GeV were used in our fitting.  That is,
only 14 of the 40 data points plotted in Fig.~\ref{fig12} were used in
the fitting. Clearly, for all $Q_T$ up to 2~GeV, the $b$-space
resummed $Q_T$ distributions are in excellent agreement with the data
from E288.  In Fig.~\ref{fig13}, we plot the resummed $Q_T$
distributions along with Fermilab E605 data \cite{DY-E605}.  The
overall normalization constant for E605 
is one.  The exactly same values of $g_1$ and $g$ are used for
calculating the theory curves in Figs.~\ref{fig12} and \ref{fig13}.
In Fig.~\ref{fig13}, from top to bottom, the four curves along with
four data sets correspond to $Q\in (7,8)$~GeV, $(8,9)$~GeV,
$(10.5,11.5)$~GeV, and $(13.5,18.0)$~GeV, respectively.  Similar to
E288 data, only two of the four data sets:  $Q\in (7,8)$~GeV and
$(10.5,11.5)$~GeV with $Q_T<1.4$~GeV were used in our fitting.
Although only 14 of the 40 data points in Fig.~\ref{fig13} were used
in the fitting, the $b$-space resummed $Q_T$ distributions are in a
good agreement with all 40 points, except a few points with $Q\in
(7,8)$~GeV.  Actually, seven of the ten data points in this set with
$Q\in (7,8)$~GeV were used in our fitting.  However, because of the
relatively large error bars, these data points did not have enough
weight in the fitting.  Nevertheless, the
theory curves calculated with two fitting parameters: $g_1$ and $g$
give a very good description of the low energy Drell-Yan data in
Figs.~\ref{fig12} and \ref{fig13}.  In particular, the overall
normalization constants for both experiments are very close to the
unity. 

As pointed out in Ref.~\cite{LBLY}, the $b$-space resummed $Q_T$
distributions have to multiply a large overall normalization constant
in order to be consistent with the Fermilab E772 data.  In
Fig.~\ref{fig14}, we plot the $b$-space resummed $Q_T$ distributions
along with the E772 data \cite{DY-E772}.  Three data sets from top to
bottom correspond to $Q\in (5,6)$~GeV, $(8,9)$~GeV, and 
$(11,12)$~GeV, respectively.  The theory curves are calculated with
the same formula and parameters used to calculate the curves in
Figs.~\ref{fig12} and \ref{fig13}, except a much larger overall
normalization constant N$_{\rm E772}=1.6$, which is consistent with
what was found in Ref.~\cite{LBLY}.  Although none of these
data points in Fig.~\ref{fig14} were used in our fitting, the
theory curves describe the data well.  Without plotting another
figure, we state that the $b$-space resummed $Q_T$ distributions are
also consistent with the R209 data \cite{DY-R209}. 

In order to get a feeling on the relative size of the contributions
from the partons' intrinsic
transverse momentum and the dynamical power corrections, we separate
the $g_2$ and $\bar{g}_2$ terms in Eq.~(\ref{g2-g}).  With
$b_{max}=0.3$~GeV$^{-1}$ and a fixed $\alpha=0.15$, we perform a new
fit with three fitting parameters: $g_1$, $g_2$, and $\bar{g_2}$.  For
the same 28 data points, we did not get much improvement in the total
$\chi^2$.  This result should be expected because (1) we already have
an excellent fit with a $\chi^2/d.o.f\sim 1$ when $g_2$ and $\bar{g}_2$
terms are combined, and (2) the range of $Q$ values in our 28 data
points are limited.  Nevertheless, the fitting result indicates that
the $g_2$ term and the $\bar{g}_2$ term are roughly equal.  That
is, the effect of partons' intrinsic transverse momentum, which is
$Q$-independent, is as important as that of the dynamical power
corrections at the fixed target energies.

From all three figures in Figs.~\ref{fig12}, \ref{fig13}, and
\ref{fig14}, we conclude that the CSS resummation formalism in
combination with our derived extrapolation provides a very good
description of the Drell-Yan $Q_T$ distributions at the fixed target
energies.  Using only 28 data points with $Q_T<1.4$~GeV to fix the 
parameters: $g_1$ and $g$, our
numerical results are consistent with over 100 data points from three
experiments.  In particular, except the E772 data, the overall
normalization constants between the theory and the data are extremely
close to the unity.

\begin{figure}
\begin{center}
\epsfig{figure=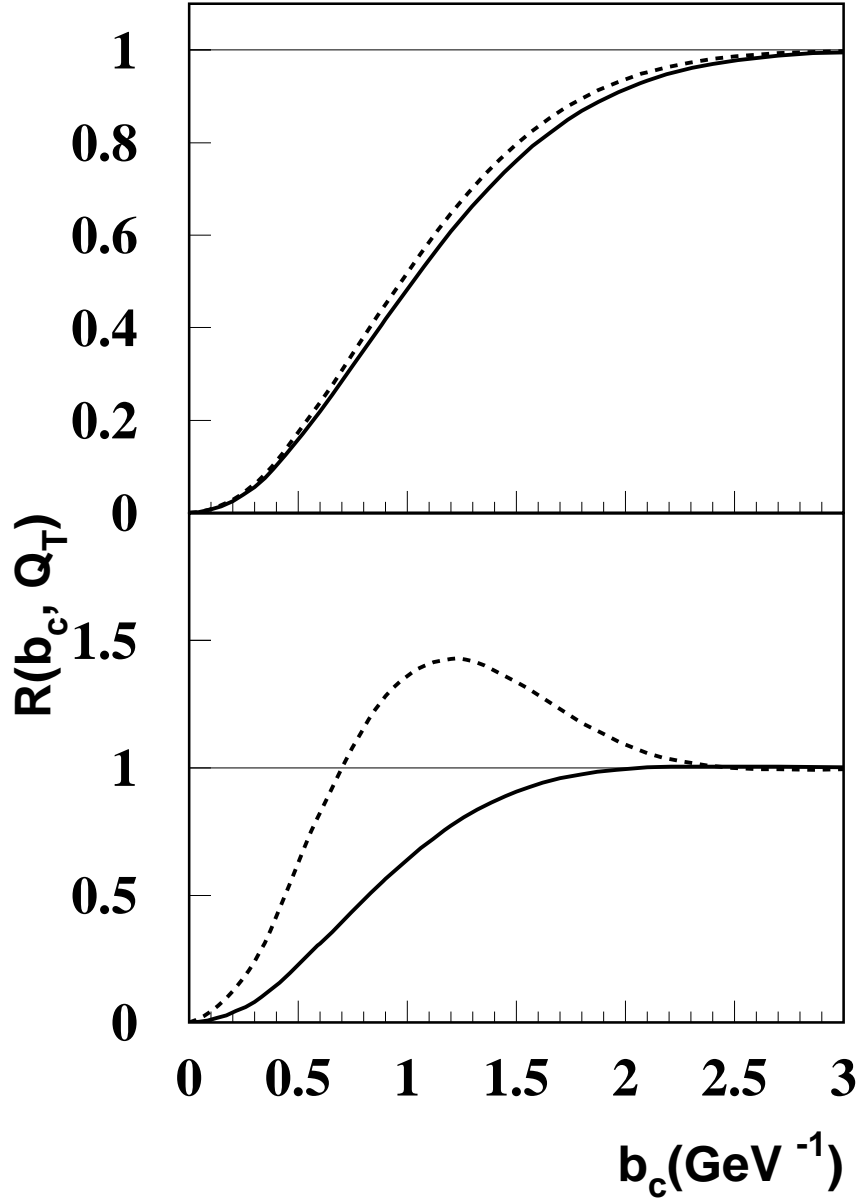,width=4.5in}
\end{center}
\caption{Ratio $R(b_c,Q_T)$ as a function of $b_c$ for Drell-Yan
production at $Q=6$~GeV and $\protect\sqrt{S}=27.4$~GeV:
$Q_T=0$~GeV (solid), $Q_T=0.5$~GeV (dashed) in the top plot;
$Q_T=1$~GeV (solid), and $Q_T=2$~GeV (dashed) in the bottom plot.}  
\label{fig15}
\end{figure}

To conclude this subsection, we demonstrate the convergence and 
stability of the Fourier transform at the fixed target energies by
plotting the ratio $R(b_c,Q_T)$ as a function of $b_c$ in
Fig.~\ref{fig15}.  The ratio $R(b_c,Q_T)$, defined in
Eq.~(\ref{W-bc}), are evaluated at $Q=6$~GeV and $\sqrt{S}=27.4$~GeV.
For calculating the resummed $b$-space distribution
$\tilde{W}^{QZ}(b,Q,x_A,x_B)$, we choose $b_{max}=0.3$~GeV$^{-1}$ and
the corresponding fitting parameters listed 
in Table~\ref{table}, which are the same as those used for plotting
theory curves in Figs.~\ref{fig12}, \ref{fig13}, and \ref{fig14}.  
In Fig.~\ref{fig15}, we plot the ratio $R(b_c,Q_T)$ for four different
$Q_T$ values: 0, 0.5, 1.0, and 2.0~GeV.  The ratio $R(b_c,Q_T)$
quickly converges to one at $b_c\sim 3$~GeV$^{-1}$.  By comparing
Fig.~\ref{fig9} and Fig.~\ref{fig15}, we conclude that the Fourier
transform for the $b$-space resummed $Q_T$ distributions converges at
$b_c\sim$~a few GeV$^{-1}$, and therefore, higher power corrections
are not very important.  On the other hand, because the Fourier
transform converges in the nonperturbative region, a reliable
extrapolation from the perturbatively calculated $b$-space
distribution at small $b$ to the nonperturbative large $b$ region is
necessary and crucial in order to be consistent with experimental data.  
From the consistency shown in Figs.~\ref{fig12}, \ref{fig13}, and
\ref{fig14}, we further conclude that the functional dependence on the
impact parameter $b$, introduced in our $F^{NP}_{QZ}$, catches the
correct physics.

\subsection{Conclusions}

In conclusion, we have quantitatively investigated the role of the
nonperturbative input in the CSS $b$-space QCD resummation formalism
for the Drell-Yan $Q_T$ distributions at both collider and fixed target
energies.  We find that the predictive power of the CSS resummation
formalism has a strong dependence on the collision energy $\sqrt{S}$.
The $\sqrt{S}$ dependence is a consequence of the steep 
evolution of parton distributions at small $x$, and it significantly
improves the predictive power of the CSS formalism at collider
energies, in particular, at the LHC energy.  We show that although the
resummed $Q_T$ distributions are mostly determined by the
perturbatively calculated $b$-space distributions at small $b$, a
reliable extrapolation to the nonperturbative large $b$ region is
necessary to ensure the correct cancellations of the $b$-integration
when $b>1/Q_T$.  By adding power corrections to the renormalization
group equations in the CSS resummation formalism, we derive a new
functional form in Eq.~(\ref{qz-W-sol-m}) to extrapolate the
perturbatively resummed $b$-space distributions to the large $b$
region.  We demonstrate that at collider energies, the CSS resummation
formalism with our extrapolation has an excellent predictive power for
$Q_T$ distributions of $W$ and $Z$ production for $Q_T$ as large as
$Q$.  Because of the smooth resummed $Q_T$ distributions, the
matching between the resummed and fixed-order calculations at large
$Q_T$ is less ambiguous. 

In this paper, we explicitly show that the power corrections are very
important for the resummed $Q_T$ distributions at the fixed target
energies.  With only two parameters: $g_1$ and $g$, obtained by
fitting 28 data points, the calculated $Q_T$ distributions using the
CSS formalism plus our extrapolation are in a good agreement with all
existing data.  In our derived extrapolation in
Eq.~(\ref{qz-W-sol-m}), the resummed $b$-space distributions for
$b<b_{max}$ do not include perturbative power corrections, while 
the nonperturbative extrapolation have both leading power
contributions as well as the power corrections.  Therefore, by
choosing a smaller $b_{max}$, we effectively move the power
corrections to the relatively small $b$ region.  We find that by
reducing $b_{max}$, the overall normalization constants for E288 and
E605 data sets are driven to the unity.  This result not only shows a
good consistency between different experiments, but also tells us that
the power corrections are very important for describing the low energy
Drell-Yan transverse momentum distributions.

Finally, we argue that the CSS $b$-space resummation formalism
should provide a reliable prediction for the Higgs production at the
LHC energy.  At the $\sqrt{S}=14$~TeV, we expect the partonic
subprocess: $g+g\rightarrow H+X$ to dominate the Higgs production when
Higgs mass $M_H\sim 115$~GeV.  From the fact that $x_A\sim x_B\sim
0.008$ are small, and gluon distribution has a steeper evolution than
quark distribution at small $x$, we expect the $\sqrt{S}$ dependence
of the $\tilde{W}(b,Q,x_A,x_B)$ to move the saddle point $b_0$ to a value
much smaller than the $b_{\rm SP}=0.37$ estimated by using
Eq.~(\ref{css-bsp}), and most likely, even smaller than 0.13 shown in
Fig.~\ref{fig3}(b) for $Z$ production at the LHC energy.  Therefore,
the $b$-space QCD resummation formalism should be valid for predicting 
the $Q_T$ distribution of hadronic production of Higgs bosons at the
LHC energy. 



\section*{Acknowledgment}

We thank P. Nadolsky and C.P. Yuan for help on the Legacy program
package, and thank S. Kuhlmann for help on experimental data.  
This work was supported in part by the U.S. Department of Energy under
Grant No. DE-FG02-87ER40731.



\begin{table}
\begin{tabular}{ccc|ccc|ccc|ccc|ccc|ccc|ccc}
  &$b_{max}$ 
  &&& $\chi^2$ 
  &&& $\alpha$ 
  &&& $g_1$ 
  &&& $g$ 
  &&& N$_{\rm E288}$ 
  &&& N$_{\rm E605}$ & \\ 
  &(GeV$^{-1}$)
  &&& \ 
  &&& \ 
  &&& (GeV$^{\alpha}$) 
  &&& (GeV$^2$) 
  &&& \ 
  &&& \ & \\ \hline
  & 0.5 
  &&& 78 
  &&& 0.65 
  &&& 0.4  
  &&& 0    
  &&& 0.85 
  &&& 0.9  & \\ \hline
  & 0.5 
  &&& 32 
  &&& 0.15 
  &&& 1.14 
  &&& 0.19 
  &&& 0.88 
  &&& 0.93 & \\ \hline
  & 0.4 
  &&& 27 
  &&& 0.15 
  &&& 1.06 
  &&& 0.23 
  &&& 0.93 
  &&& 0.98 & \\ \hline
  & 0.3 
  &&& 24 
  &&& 0.15 
  &&& 0.92 
  &&& 0.28 
  &&& 0.97 
  &&& 1.00 &
\end{tabular}
\vskip 0.2in

\caption{Nonperturbative parameters in the $F^{NP}_{QZ}$ obtained by
fitting 28 data points on Drell-Yan $Q_T$ distributions at the fixed 
target energies.}
\label{table} 
\end{table}

\end{document}